# Insights into the defect-driven heterogeneous structural evolution of Ni-rich layered cathode in lithium-ion batteries†

Zhongyuan Huang,[a] Ziwei Chen,[a] Maolin Yang,[a] Mihai Chu,[b] Zenan Li,[c] Sihao Deng,[de] Lunhua He,[efg] Lei Jin,[h] Rafal E. Dunin-Borkowski,[h] Rui Wang,*[i] Jun Wang,*[c] Tingting Yang,*[h] and Yinguo Xiao*[a]

[a] School of Advanced Materials, Peking University, Shenzhen Graduate School, Shenzhen 518055, PR China
[b] Department of Chemistry, Materials, and Chemical Engineering "Giulio Natta", Politecnico di Milano, Milano 20133, Italy
[c] School of Innovation and Entrepreneurship, Southern University of Science and Technology, Shenzhen 518055, PR China
[d] Institute of High Energy Physics, Chinese Academy of Sciences, Beijing 100049, PR China
[e] Spallation Neutron Source Science Center, Dongguan 523803, PR China
[f] Beijing National Laboratory for Condensed Matter Physics, Institute of Physics, Chinese Academy of Sciences, Beijing 100190, PR China
[g] Songshan Lake Materials Laboratory, Dongguan 523808, PR China
[h] Ernst Ruska-Centre for Microscopy and Spectroscopy with Electrons, Forschungszentrum Jülich GmbH, Jülich 52428, Germany
[i] Department of Engineering, University of Cambridge, Cambridge CB30FS, UK

* Corresponding Authors: Jun Wang (wangj9@sustech.edu.cn), Rui Wang (rw716@cam.ac.uk), Tingting Yang (t.yang@fz-juelich.de), Yinguo Xiao (y.xiao@pku.edu.cn)

† Electronic supplementary information (ESI) available.

## Abstract

Recently, considerable efforts have been made on research and improvement for Ni-rich lithium-ion batteries to meet the demand from vehicles and grid-level large-scale energy storage. Development of next-generation high-performance lithium-ion batteries requires a comprehensive understanding on the underlying electrochemical mechanisms associated with its structural evolution. In this work, advanced *operando* neutron diffraction and four-dimensional scanning transmission electron microscopy techniques are applied to clarify the structural evolution of electrodes in two distinct full cells with identical $LiNi_{0.8}Co_{0.1}Mn_{0.1}O_2$ cathode but different anode counterparts. It is found that both of cathodes in two cells exhibit non-intrinsic two-phase-like behavior at the early charge stage, indicating selective $Li^+$ extraction from cathodes. But the heterogeneous evolution of cathode is less serious with graphite-silicon blended anode than that with graphite anode due to the different delithiation rate. Moreover, it is revealed that the formation of heterogeneous structure is led by the distribution of defects including Li/Ni disordering and microcracks, which should be inhibited by assembling appropriate anode to avoid potential threaten on cell performance. The present work unveils the origin of inhomogeneity in Ni-rich lithium-ion batteries and highlights the significance of kinetics control in electrodes for batteries with higher capacity and longer life.

## Introduction

Developing environmentally friendly and renewable energy storage techniques is one of the key goals to reduce carbon emissions. Among all energy storage devices, lithium-ion battery (LIB) is one of the most widely used devices due to the economic advantages, long life, and high capacity.[1] Nowadays, to satisfy the requirements of higher energy density and longer cycling life for batteries, Ni-rich layered cathode[2] ($LiNi_xCo_yMn_{1-x-y}O_2$, x > 0.5) and graphite-silicon composite[3] anode are applied in the new-generation power battery pack for electric vehicles. However, compared to layered $LiCoO_2$ or $LiNi_{0.33}Co_{0.33}Mn_{0.33}O_2$, Ni-rich cathodes, such as $LiNi_{0.8}Co_{0.1}Mn_{0.1}O_2$ (NCM811), show a lower initial Coulombic efficiency (CE) and quicker capacity fade during cycling, implying a more serious and irreversible structural degeneration process.[2,4] Hence, careful preparation and modification are necessary for Ni-rich layered oxides to improve the electrochemical performance.[5] As for anodes, especially silicon, although its theoretical capacity (3572 mAh $g^{-1}$, to $Li_{15}Si_4$) is nearly 10 times than that of commercial graphite (372 mAh $g^{-1}$, to $LiC_6$), the serious particle expansion and pulverization during lithiation restrict the further commercialization.[3,6] Up to now, an alternative approach is to mix slight silicon and/or silica with graphite to balance the capacity and cycling life of the composite anode.

Despite the enhancement of electrochemical performance, investigations on LIB full cell systems with complex couplings among components and scales is still a big challenge,[7] especially for the large-size devices in application.[8] Different from coin cells or simulated cells in laboratory, heterogeneous charge/discharge process in practical energy storage cells is more common and pronounced,[9-12] which could reduce the cell capacity and hinder the cycling stability. Systematic understanding of the multi-scale heterogeneity of electrodes during cycling requires the combination of multiple characterizations. Previous works had developed several *operando* characterization methods, such as X-ray diffraction (XRD),[13-15] Raman spectroscopy,[7] X-ray absorption spectroscopy (XAS),[15] online electrochemical mass spectrometry (OEMS),[16] etc. to investigate the internal electrochemical mechanisms of charged cells. Although many interesting phenomena were found, explained, and reported, it is still insufficient to reach a consensus on inducements of heterogeneous electrochemical process in full cells.

Herein, we propose the utilization two (or more) complementary characterization methods to investigate this issue from both the global and local view. First, *operando* neutron diffraction experiments were considered to collect the overall structural information of cathode, anode, and other parts of cells during electrochemical process.[17-19] As a unique probe for microstructure, neutrons possess stronger penetrability than X-ray and electrons, allowing them to pass through a thick pouch cell with information from internal positions. Additionally, neutron is mild to materials, which would not cause serious damage or influence the cycling of cell during the experiment.[20] As for the elements in a LIB, like Li, Ni, and C, the neutron coherent scattering lengths ($b_c$, analogue of the X-ray form factor $f$) have quite different values (-1.90, 10.3, and 6.65 fm respectively), which not only guarantees the sensitivity to Li, but also permits the synchronous observations of cathode and anode structure. Moreover, $b_c$ could not decay with wave transfer vector $Q$ ($Q = 2\pi/d$.) increasing, which enables more reliable reflection signals for more accurate structural information determination, such as vacancy[21] or cation disordering defects.[22] In brief, the neutron probe has its advantages in particular studies of full cells.

Next, to compensate the spatial resolution under particle scale, four-dimensional scanning transmission electron microscopy (4D-STEM)[23] was used to record and map the structural state of electrode particles from the local perspective. With this technique, a focused electron beam is scanned over the sample, and an electron diffraction is recorded for each pixel position. The resulting data can be utilized for various quantitative analysis methodologies to measure the sample properties with nanometer resolution. 4D-STEM allows for versatile data acquisition and precise structural analysis. It overcomes some limitations of atomic resolution imaging techniques, particularly in terms of reducing electron dose, enabling larger fields of view, etc. Especially, precession-assisted 4D-STEM measurement could minimize the effects of dynamical electron interactions while maintaining the geometry of the diffraction pattern, and enhance the imaging of higher order reflections, by rocking Ewald's sphere through more of the reciprocal space. As a result, information from electron diffraction patterns is significantly enhanced.

In this work, detailed structural evolution process in full cells was investigated according to the above ideas. To get close to the practical case, we selected two representative cell systems for current study. NCM811 was used as cathodes for both two cells, while the anodes are different for these two cells. One was carried with graphite anode, another assembled with composite anode with graphite and minor silicon-silica mixture, labeled as GSO. Their electrochemical performances show that NCM811 exhibited a more stable cycling behavior with composite anode, while a low initial discharge capacity followed by a continuous capacity increase was observed in the cell with pure graphite. Through the combined of *operando* neutron diffraction and 4D-STEM, a non-uniform structural evolution was revealed at the early charge state, which is due to the difference in reaction activity or delithiation ability among particles. However, the relative slow kinetics of composite anode mitigated the phase heterogeneity and reached the equilibrium state during the $1^{st}$ charging, while the cell with graphite anode could not eliminate the phase heterogeneity during the $1^{st}$ cycle, leading to a non-uniform lithium distribution in primary particles of cathode and a lower initial capacity. Further analysis indicates that the evolution of local defects with $Li^+$ extraction would influence the further delithiation of particles. $Li^+$ located in the region with less Li/Ni disordering is preferable to migrate firstly. And the strain accumulation with $Li^+$ extraction causes the formation, growth, and penetration of microcracks, which provides more interfaces to induce more $Li^+$ migration. The evolution mechanisms of cathode were controlled by the kinetics of cells. A relative slow delithiation process would reduce the heterogeneity phenomenon and build a more stable and long-life cell system. Our findings unveil the kinetic effect on the electrochemical process and point out the importance of electrode design to cell performance.

## Results

### *Operando* neutron diffraction studies for LIB full cells

Schematics of *operando* neutron diffraction experimental set-up and the full cell design are shown in Fig. 1a and Fig. 1b. Cells were fixed in the scattering cavity and connected to the external device for electrochemical measurements. Considering the balance between the counting rate and data acquisition time, the data collection plan was finally determined as that in Fig. S1, ESI†. Once started, proportion of incident neutrons were scattered and collected by the detector banks. Simultaneously, structural information and evolution of electrodes in the dynamic process were recorded by scattered neutron signal. It should be noted that the neutron data collection time is still longer than that of X-ray data due to the relative lower flux of neutron beams, although facility was gradually optimized. Hence, collected *operando* patterns virtually reflect the time-averaged structure from a series of dynamic evolution traces during that period. Construction of full cells for *operando* experiments is schematic illustrated in Fig. 1b. Maximized beam size in the experiments loses the spatial resolution to cells. The spatial distribution evolution of lithium in electrodes is unable to investigate under this condition. To compensate that, *ex-situ* 4D-STEM was applied for primary particles of charged cathodes, which would be introduced and discussed in corresponding sections.

Through *operando* experiments, states of voltage and the electrode structures during charging/discharging are correlated by the time, as shown in Fig. 1c. To better describe and compare the electrode structures from different cells under the same progress of charging, state of charge (SOC) was proposed to normalize the time for the charge/discharge curve. Fig. 1c takes the 811|G as an example. Selected representative neutron diffraction contour plots on the right are also labelled by SOC, of which the $d$-spacing is selected within 2.38~2.52 Å and 3.22~3.8 Å. Within the former range, only (101) reflection of NCM811 is located, while only (00$L$) reflections of graphite/graphite intercalation compounds (GICs) existed within the latter range. Therefore, both the cathode and anode structural state within each charge/discharge stage could be qualitatively determined separately through the evolution of characteristic reflections in *operando* patterns. From Fig. 1c, a shift to lower $d$-spacing with split was observed for the cathode reflection during charging, but split became less-pronounced within discharging. As for the graphite anode, the (002) reflection at 3.35 Å slightly shifted to the higher $d$-spacing in the early charge stage. Then a series of GICs reflections, with fixed positions in $d$-spacing from 3.43 to 3.67 Å, appeared and declined sequentially. This reflects the staging mechanism of graphite with Li intercalation.[24,25] At the end of discharge, anode was unable to recover to pure graphite, which suggested the lithium loss during initial cycling. Details of the electrode structural evolutions and the difference in two cells are shown in the following sections.

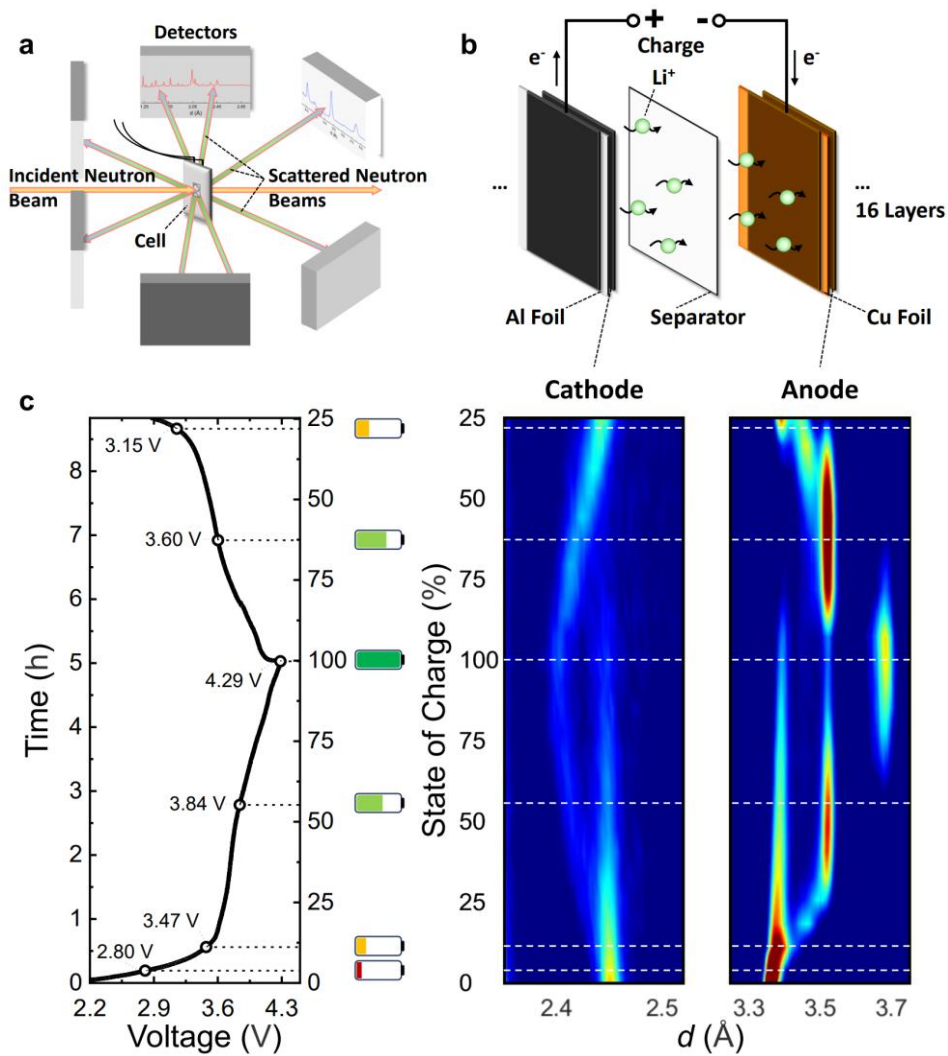

**Fig. 1** Schematic illustration of the LIB full cell *operando* neutron studies. (a) Schematic of the *operando* neutron diffraction experiments at GPPD@CSNS. (b) Structure of full cells with NCM811 cathode and graphite/GSO anode. (c) the 1st 0.2C cycle voltage-time/SOC electrochemical curve of the 811|G cell and the corresponding selected contour plots of *operando* neutron diffraction patterns containing evolutions of cathode within 2.38~2.52 Å and graphite anode within 3.22~3.8 Å.

**Fundamental information of electrodes and cells**

SEM images and EDS mappings for graphite and GSO anodes are displayed in Fig. 2a-b. For commercial graphite in Fig. 2a, more small flakes and gaps are observed in the SEM image, corresponding to the larger full width at half maxima (FWHM) of graphite reflections in neutron patterns for 811|G compared to that of 811|GSO. The smaller graphite flakes lead to a large surface area, enhancing the ion exchange with electrolyte and improving the Li$^+$ intercalation/deintercalation kinetics. As for GSO, more integral particles are observed in Fig. 2b. EDS mappings of C, O and Si clearly distinguish between the graphite and silicon-silica particles. Morphology image and EDS mappings of NCM811 cathode are also displayed in Fig. S2, ESI†. The cathode consists of micron-sized sphere-like NCM811 secondary particles connected with conductive agent/binder (Fig. S2a, ESI†). NCM811 primary particles were in irregular outlines with sizes in several hundred nanometers, which are shown in the 4D-STEM section. The EDS mapping of C (Fig. S2b, ESI†) indicates the distribution of acetylene black/PVDF, which are coated on the surface or fill the space among particles. EDS mappings of O, Ni, Co, and Mn (Fig. S2c-f, ESI†) reveal the uniform element distribution in prepared NCM811 particles.

Electrochemical measurement results for the two types of cells are shown in Fig. S3, ESI†. It is interesting that in both rate and cycling tests, 811|G displayed gradual capacity growth, or slow cell activation in the first few cycles. The normalized capacity increased from 0.75 and 0.56 to 1.00 and 0.77 separately in the first 5 cycles at 0.1C and 0.5C, while the capacity of 811|GSO was slowly decayed from the 1$^{st}$ cycle as the usual half or full cells using NCM811 as the cathode. Moreover, capacity of 811|G showed fluctuations in the following cycles. The abnormal low initial capacity, the long activation cycles, and unstable capacity of 811|G indicate that a non-equilibrium electrochemical process occurred during the initial cycling. Since 811|G both displayed similar activation behaviors in rate and cycling performance test, it is reasonable to think that this phenomenon is not an occasional individual issue, but exists generally in this cell system. The *operando* neutron diffraction experiments supply abundant structural information for further understanding what had happened in cells during the 1$^{st}$ cycle.

Charge-discharge curves and corresponding differential capacity curves of two cells in the *operando* experiments are shown in Fig. S4, ESI†. And the contour plots for *operando* neutron patterns of 811|G and 811|GSO are displayed in Fig. 2c-d. Since both anodes were not pre-lithiated before assembly, an initial open-circuit voltage (OCV) of near 0 V and a notable lithium loss after the 1$^{st}$ cycling could be observed. From the curves in Fig. S4a, ESI†, recorded during neutron studies, the two cells had a low initial CE (81.2% for 811|G and 80.5% for 811|GSO), resulting from solid electrolyte interphase (SEI) formation, interface side reactions and other irreversible lithium loss as predicted. The 811|G displayed a lower initial cathode discharge specific capacity of 127.7 mAh g$^{-1}$ than 811|GSO of 147.9 mAh g$^{-1}$, corresponding to the abnormal poor electrochemical performance observed in the previous electrochemical measurements. Two cells had similar CE but different level of capacity of the 1$^{st}$ cycle, implying less amount of Li extracting from NCM811 in the 811|G cell during 1$^{st}$ charging. To further understand the difference of electrochemical process in two cells, dQ dV$^{-1}$ curves of the initial cycle are plotted in Fig. S4b, ESI†. Generally, a peak in the dQ dV$^{-1}$ curve represents a platform in the electrochemical curve, which usually corresponds to a phase transition process in electrodes. However, for NCM811 cathode, previous reports using very high-resolution synchrotron X-ray diffraction did not show a LiNiO$_2$-like discrete evolution[26,27] of reflection peaks under a low rate during cycling.[14,15] This fact suggests that the continuous solid-solution behavior should be the intrinsic evolution process of NCM811. Therefore, phase notations, like H1, M, H2, etc., are not applied in this work. Only in Fig. S4b, ESI†, the so-called "phase transitions" are given, which are just used to identify peaks observed in dQ dV$^{-1}$ curves, not representing an observed phase transformation from *operando* neutron data. From the labelled peaks, it is found that the overall profiles of 811|G and 811|GSO are similar, except peak split for 811|G and the incomplete peak of the so-called "H2→H3 transition" at high voltage for 811|GSO during charging. It is deduced that in 811|G, cathode structure might not evolve homogeneously, resulting in an asynchronous phase evolution. And the cathode structural evolution in 811|GSO should lag that in 811|G.

Structural information from the *operando* neutron diffraction contour plots in Fig. 2c-d could confirm the deduction above. Although the very complex reflection distribution is shown from 0.5 to 2.52 Å in the plots, it is not difficult to recognize the attribution of diffraction peaks. Since *c*-axis of graphite expands continuously during lithiation, the 00*L* reflection will shift to high *d*-spacing range.[28] The increased valence of transition metal (TM) ions, or strengthened interaction between TM and neighboring oxygen during delithiation,[29,30] will lead to the shrinkage of *ab*-planes in layered cathodes, resulting in the shift of 10*L* reflections to low *d*-spacing range. As for current collectors or separators, no shift will be observed during the whole process. Based on the properties above, origins of each reflection in the contour plots are found. Some typical reflections from different components of cells are labelled in Fig. 2c-d, part of which are indexed according to the *d*-spacing. The two *d*-spacing ranges of 1.65~1.9 Å and 2.37~2.49 Å in white dash boxes of Fig. 2c correspond to the patterns displayed in Fig. 1c. The interval of 1.65~1.9 Å, which is near half of 3.22~3.8 Å, covers graphite/GICs (00 2*L*) reflections with the same evolution as (00*L*) reflections. Based on the contour plots, the overall structural evolutions of anode and cathode in two cells are clearly exhibited.

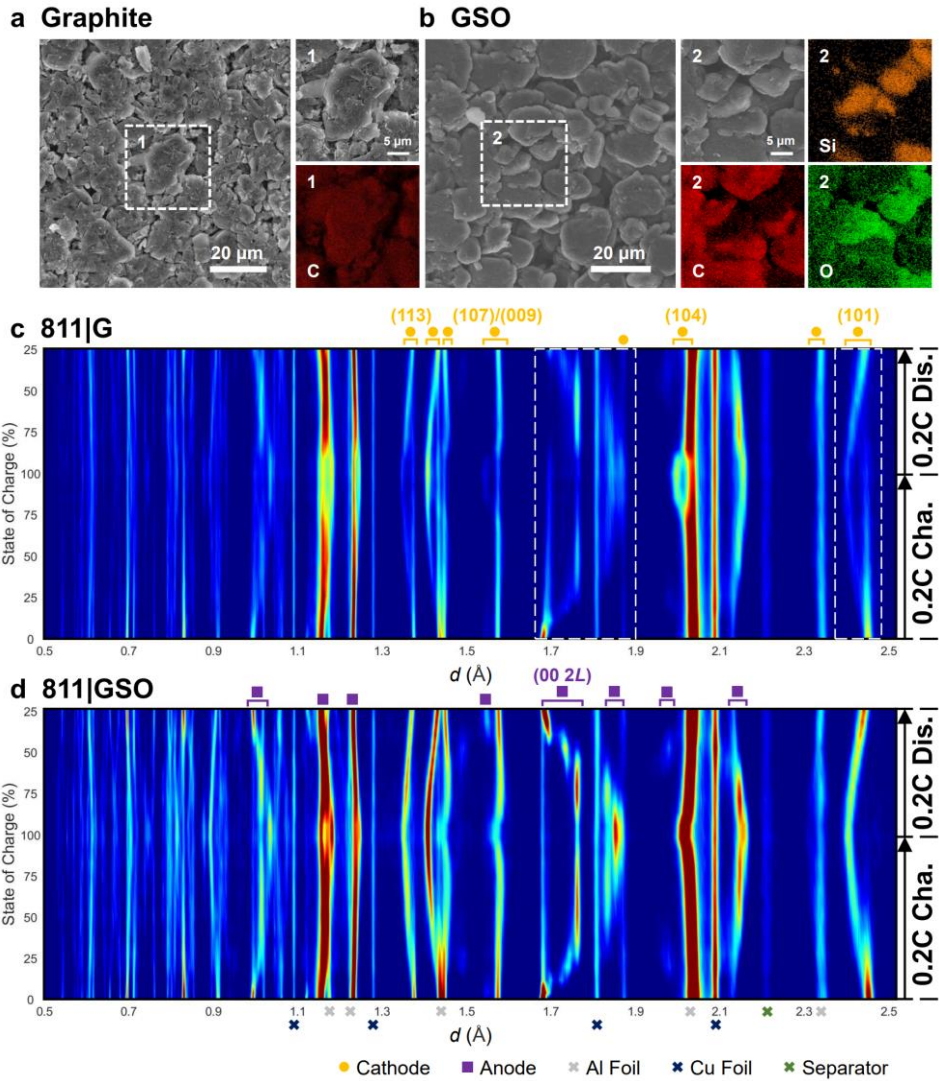

**Fig. 2** Difference in anodes and corresponding *operando* neutron diffraction results. (a) graphite and (b) GSO anode SEM morphology images. Enlarged images of framed area 1 in (a) and area 2 in (b) with EDS mappings (carbon for area 1, carbon, silicon, and oxygen for area 2) are attached to corresponding figures. Contour plots of (c) 811|G and (d) 811|GSO during the 1st 0.2C cycling over the range from 0.5 to 2.52 Å. Contributions of NCM811, lithiated graphite, current collectors, and separators to the diffraction patterns are marked, among which representative reflections are indexed. Evolutions of NCM811 (101) reflections and lithiated graphite (00 2*L*) reflections during charging/discharging are framed by white dash lines.

**Structural evolutions of graphite anodes**

For anodes, due to the low content and poor crystallinity of silicon and silica, it was unable to be observed in *operando* patterns of Fig. 2d. Therefore, we focus on the phase evolution of lithiated graphite phases. Li$^+$ intercalation in graphite induces staging transitions[25] with discrete reflection evolutions. Several separated reflections appeared sequentially near the graphite (004) reflection within 1.65~1.9 Å in Fig. 2c-d with the increase of SOC. Each of them represented one GIC species[31] with different Li content, which is summarized in Fig. S5a, ESI†. Diffraction patterns within 3.25~3.75 Å in Fig. S6, ESI†, are used to illustrate the transition process, since no interference reflection from other cell components locates in this region. At the very beginning of lithiation, which is marked as "I", lithiated graphite maintains its original structure, known as the 2H phase of space group (S.G.) $P6_3/mmc$. The (002) reflection only displayed a slight expansion without broadening or splitting, indicating no change in stacking sequence of graphene layers and random $Li^+$ distribution in the highly-diluted lithiated graphite. In previous results, the structure state with ultra-low lithium content is usually denoted as the stage 1L or "gas-like" phase.[25] After a short period for stage 1L, more reflections appeared in the patterns, with significant overlap from 3.37~3.51 Å. FWHM of them are larger than that of the original graphite, suggesting the change of graphite layer stacking. During the period marked as "II" in Fig. S6, ESI†, several GIC species with different $Li^+$ arrangements in graphite layers were formed, such as $LiC_{18}$, $LiC_{30}$, and $LiC_{40}$, or denoted as stage 2L, 3L, and 4L. However, the duration of existence for each phase is close to the data acquisition time for one *operando* pattern. It is difficult to extract more details from these GICs. We only claim that there were several $LiC_x$ (x > 12) species appearing sequentially during the period "II", resulting in significant phase heterogeneity. After that, two strong reflections corresponding to the well-crystallized phases $LiC_{12}$ and $LiC_6$ (S.G. $P6/mmm$ for both), also called stage 2 and stage 1, were observed during periods "III" and "IV", respectively, in Fig. S6, ESI†. The charge process stopped during the transition from $LiC_{12}$ to $LiC_6$. Because of the excessive anode amount, graphite cannot be fully lithiated when charged to 100% SOC. During discharging, inverse process occurred sequentially, roughly analogous to those in the charging process. Positions of (00*L*) reflections and the attributions are labelled in Fig. S6a, ESI†.

Comparing the graphite transition in the two cells, it is observed that graphite did not alter its intrinsic mechanism, but some differences were found between the two anodes. First is the evolution of residual dilute lithiated graphite during cycling. After stage 1L, dilute phase started to transform into several GICs like $LiC_{40}$, $LiC_{30}$, etc., as above described. However, according to *operando* patterns in Fig. S6, ESI†, not all dilute phase was consumed. Some maintained the graphite-like structure and was gradually reduced with further charging. The heterogeneous phase evolution is probably due to the non-uniform distribution of charging current,[12] electrolyte,[10] electrode thickness,[32] etc. As for 811|G, the continuous $Li^+$ diffusion into the graphite promoted the evolution of residual dilute phase, resulting in the weakened reflections and the shift towards higher *d*-spacing range (Corresponding to the stage 1L→stage 2L/3L/4L transition). However, the dilute phase reduction in 811|GSO was slower. Due to the much higher specific capacity and the relatively lower potential (versus Li) of silicon/silica than graphite,[33] silicon-based part of GSO contributed more capacity during the early lithiation stage,[34] leading to less $Li^+$ intercalation on the graphite part. With less $Li^+$ supply, less dilute phase was transformed to GICs, resulting in the stage 1L reflections maintained during the whole cycle.

The GICs in the two cells also displayed different evolution pathways during lithiation. In 811|G, multiple reflections corresponding to high-order GICs (GICs except $LiC_{12}$ and $LiC_6$) within 3.4~3.5 Å were observed when SOC > 18.5%. When $LiC_{12}$ appeared, reflections of other GICs decreased to a low level, but they were not eliminated at higher SOC. A continuous diffraction "band" between the reflections of dilute phase and $LiC_{12}$, which is obviously higher than the background signal, was detected when SOC > 33.3%. This phenomenon is also observed in other works,[35,36] indicating multiple GIC phase distribution induced by lithium concentration gradient. In 811|GSO, the gradient structure was less-pronounced, with only weak signal close to either dilute phase or $LiC_{12}$ for SOC > 35.3%. The relative more uniform phase distribution of GICs in 811|GSO could be accounted by the slower $Li^+$ intercalation into the anode. Based on previous reports,[12,36,37] when the lithium intercalation rate on the surface is greater than the lithium diffusion rate within the graphite particle, more lithium will concentrate on the edges of graphite flakes, resulting in the concentration gradient from the surface to the bulk. Therefore, gradient structure of GIC phase formed in 811|G under the condition of insufficient lithium diffusion. In 811|GSO, since silicon-based particles dominated the reaction within the early charging stage, less $Li^+$ migrated to graphite particles for insertion. Consequently, $Li^+$ could diffuse fully in graphite, leading to the reduced gradient structure for graphite in 811|GSO, as observed in Fig. 2d and Fig. S6b, ESI†.

The last difference is the final structural state of graphite in two cells. In 811|G, dilute lithiated graphite coexisted with $LiC_x$ (x > 12) and residual $LiC_{12}$, indicating that not all $Li^+$ were extracted from the graphite. Considering the $Li^+$ for SEI formation, lithium loss was estimated to be serious after the 1$^{st}$ $Li^+$ intercalation/deintercalation. However, the graphite in 811|GSO returned to almost the initial structure according to Fig. S6b, ESI†. But it did not suggest the highly reversible structural evolution. Since graphite was lithiated later than silicon/silica during charging, it should also delithiate prior to that during discharging. Therefore, we infer that there was still $Li^+$ remaining in GSO, mainly including the $Li^+$ in silicon-based particles and that in the SEI film, resulting in a certain degree of lithium loss.

**Phase heterogeneity in cathodes**

Although the different anode composition, great discrepancy in evolutions of cathode structure is observed by the *operando* experiments. Before charge, NCM811 cathode has a rhombohedral layered structure (S.G. $R\bar{3}m$, but mostly using hexagonal axes). After charge start for a shot period, reflection split emerged from (101) at ~2.44 Å and (113) at ~1.37 Å in both cells. In 811|G, split lasted for a long period (SOC from 33.33%/~3.67 V during charging to 55.56%/~3.52 V during discharging) in the 1$^{st}$ cycle, while it gradually vanished in 811|GSO (SOC from 27.41%/~3.79 V to 74.41%/~4.08 V during charging). Split in Ni-rich layered cathode family usually implies two possible mechanisms.[38] One is the reduced symmetry of the crystal lattice from the rhombohedral to the monoclinic, known as the reported H1→M transition at relative low voltage in LiNiO$_2$-like cathodes.[26,27,39] As for that in this work, to the best of our knowledge, we think it should not be attributed to the phase transition based on the following reasons: First, as discussed above, previous works have found that the delithiation of polycrystalline NCM811 follows the solid-solution mechanism, with no phase transition like that in LiNiO$_2$ being observed during cycling.[14,15] Second, a common H1→M transition usually finishes in the low-voltage stage, because the internal mechanisms, known as the Jahn–Teller distortion of Ni (III) or Li-vacancy ordering, both require the low degree of delithiation.[38] Consequently, the transition-induced reflection spilt cannot exist at the high-delithiation state. Third, the transition is reversible according to the dQ dV$^{-1}$ curves in Fig. S4b, ESI†, indicating that split should also occurred during discharging, if it is really due to the structural transition. However, no reflection split is observed in the discharge patterns of 811|GSO in Fig. 2d. Therefore, the split observed in this work is less possibly originated from phase transition.

Another possibility taken into consideration is that it is not the intrinsic phase evolution process of NCM811, i.e., a kinetic-induced two-phase-like behavior is accounted for the split of reflections, which is mostly reported under fast charging conditions.[40,41] But some studies have also pointed out that the two-phase-like phenomenon could also be found in the cases without kinetic limitations, e.g., at low charging rates or in a relaxation state.[42,43] In this work, the latter case is in line with our observation. At the beginning, due to the heterogeneous reactivity among cathode particles resulting from various reasons (surface composition, size, location on the cathode, etc.),[44-46] partial particles were preferentially activated. With the continuous electrochemical reaction of the cathode, lattice parameters of partial NCM811 gradually displayed large volume changes, while the rest part maintained the nearly original lattice size. With the difference of lattice parameters being large enough, the two-phase-like behavior was observed by neutron diffraction. Therefore, phase heterogeneity in cathodes emerged during charging, corresponding to the observation in dQ dV$^{-1}$ curves. The variations of (101) reflections in two cells during the 1$^{st}$ cycling and some selected patterns under certain SOC are shown individually in Fig. S7, ESI†. Single (101) reflection was observed at the initial state for both cells. With the increase of SOC, the reflection broadened and behaved as a bimodal distribution, which could be simplified by a fictitious "two-phase" model. Note that "bimodal" does not represent only two separated *d*-spacing values, but a continuous distribution of *d*-spacing with two characteristic peak values within certain range, the real structure should contain a population with *d*-spacing located near one value, a second population with another *d*-spacing, and a few populations with intermediate *d*-spacing values. However, considering the majority *d*-spacings near the two peak values, it is appropriate to replace the distribution by two phases with different lattice parameters. In this work, we suggest that the phase with large lattice variation, i.e., the normally evolved part of cathode, could be defined as the kinetic active phase ($H_{act}$, where H represents the hexagonal cell). Also, the phase did not have an obvious lattice change would be defined as the kinetic sluggish phase ($H_{slu}$). Contributions of $H_{act}$ and $H_{slu}$ are marked in Fig. S7, ESI†. Except shifts of reflections, another phenomenon is the change of intensity, which indicates the transformation from $H_{slu}$ to $H_{act}$ during cycling. That is to say, the kinetic sluggish region was able to be activated as the cycle progressed. And for the two cells, 811|GSO was more favorable to finish the cathode activation. The reason will be discussed later. In summary, the active phase in NCM811 experienced the delithiation process with significant changes in lattice parameters, while the sluggish phase kept nearly unchanged during cycling. To understand the internal mechanism behind the abnormal evolution process, hidden structural information in *operando* patterns are essential. During the section of graphite anodes, previous description has shown that the multi-phase structure is due to the limited lithium diffusion within particles. Considering the difference lithium content of GICs, it could be deduced that the two-phase-like phenomenon of cathode is also originated from the different delithiation ability. To verify that, Rietveld refinements were performed to extract structural information from cathodes.

Structural models used for refinements are shown in Table S1, ESI†, which contains most crystallized components in full cells. Initial values of structural parameters were from the neutron powder diffraction results in previous works.[22,25] Refinement results of lithium content and lattice parameters with corresponding SOC are exhibited in Fig. 3. Corresponding structural evolution process is illustrated in Fig. S5b, ESI†. Refined patterns of initial states and states at the end of charge for 811|G and 811|GSO are shown in Fig. S8, ESI†. In the refined patterns, used structural models well described most neutron reflections from full cells in working condition, which verified the analysis above.

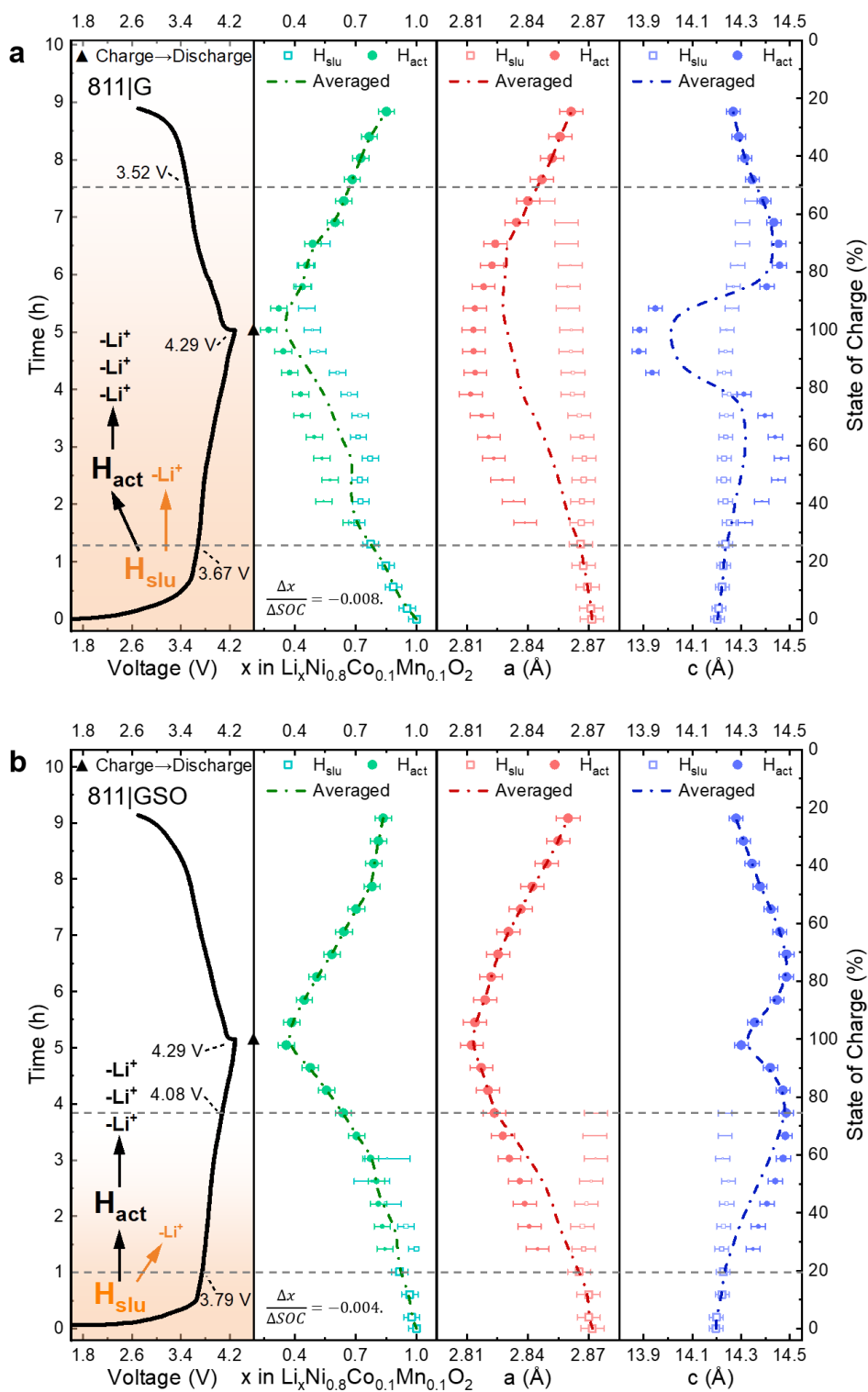

**Fig. 3** Rietveld refinement results of *operando* neutron diffraction patterns for (a) 811|G and (b) 811|GSO, including the 1st 0.2C cycling time-voltage electrochemical curve and the corresponding evolution of refined parameters (lithium content, lattice parameters *a* and *c*) with voltage marked at critical states (charge to discharge, visibility and invisibility of the two-phase behavior) during the *operando* neutron diffraction experiments. Averaged value of parameters

in the two-phase region was estimated by weighting values from each phase by the intensity of (101) reflections. Evolution processes of cathodes are simply summarized in each figure.

Initial cell parameters (*a* and *c*) of cathode were 2.872(6) Å and 14.20(3) Å for both 811|G and 811|GSO as shown in Table S2, ESI†, and Table S3, ESI†, which is close to those from powder sample in our previous work.[22] Therefore, lithium site occupancy of initial-state NCM811 in two cells was set as 1. After the 1st cycling, lithium content, lengths of *a*-axis and *c*-axis were 0.86(4), 2.863(6) Å, and 14.26(3) Å for 811|G, and 0.87(4), 2.862(6) Å, and 14.25(3) Å for 811|GSO. At the end of discharge, the cathode structure cannot recover to the initial state due to the kinetic limit of lithiation at the near full-filled state.[30] About 13~14% of the lithium did not insert back into cathode during the first cycle, which remained in the counter electrode or SEI films. The lost lithium, as one of the reasons for poor initial CE, could be reused to some extent by a following potentiostatic hold.[30] Refinement results of the initial and final states illustrated that cathode structural evolution in two cells had close initial and final states.

For the evolution process of cathode in 811|G displayed in Fig. 3a, it is obvious that the two-phase-like phenomenon emerged since the voltage was over 3.67 V. From the electrochemical curve, the cell was just at the early stage of plateau region. A similar result was also found in Fig. 3b, but the voltage point is 3.79 V. The higher voltage plateau of 811|GSO could be attributed to the larger polarization of GSO anode, reflecting slower kinetic process. Besides, more differences were found from Fig. 3a, which could unveil the internal mechanism of two-phase-like behaviors in cathode and the way for inhibition.

Before the two-phase-like stage, it was found that the lattice parameters displayed a delayed response to the delithiation. Whatever the cathode in 811|G or 811|GSO with different delithiation speeds (linear decrease of lithium content with ~0.008 Li per SOC for 811|G, and ~0.004 Li per SOC for 811|GSO from refinement results) during the early delithiation, variations of lattice parameters were smaller compared to those during further delithiation stages (*a* from 2.872(6) to 2.866(6) Å, *c* from 14.20(3) to 14.24(3) Å for 811|G; *a* from 2.872(6) to 2.865(6) Å, *c* from 14.20(3) to 14.23(3) Å for 811|GSO). This phenomenon has been reported previously.[18,47] Herein, we proposed that the sluggish evolutions of *a* and *c* were the early symptom of two-phase-like evolutions. That is, phase heterogeneity occurred at the very early charging stage due to different conditions of each particle. But at the beginning, differences in lithium content and lattice parameters between the kinetic active and the kinetic sluggish part were too small to be distinguished by neutron beams, resulting in a pseudo-homogeneous phase evolution observed by neutron diffraction. Since most region of cathode had not been activated at early charging states, the overall phase structural evolutions during this stage were close to the sluggish part, showing low volume changes in response to delithiation.

After the single-phase-like region, significant differences of lithium content and lattice parameters were observed between two phases in Fig. 3a. Average values were also be plotted by dash lines, which represented the overall structure state for the heterogeneous systems. It was found that $H_{act}$ phases in both cells shown a significant smaller *a*-axis and larger *c*-axis than the average value as it was observed, indicating the delithiation in $H_{act}$ was accumulated to a notable high degree. With lithium deintercalation from cathode particles, *a*-axis decreased gradually due to the decreased bond length between TM and oxygen until the oxidation of lattice oxygen.[27] *c*-axis expanded first due to the enhanced electrostatic repulsion between neighboring oxygen layers with delithiation, and then collapsed due to the charge transfer from O 2p to partially-filled Ni $e_g$ orbitals, which exhausted the negative charge of O.[29] Based on the refined lattice parameters, it was found that evolutions of $H_{act}$ followed the process of normal cases. In 811|G, *c*-axis of $H_{act}$ reached the maximum of 14.46(3) Å at the SOC of 55.56%, while it was 14.48(3) Å at 74.41% SOC in 811|GSO. The observed earlier *c*-axis maximum reflected the more serious lithium extraction of $H_{act}$ in 811|G. As for $H_{slu}$ phases with poor kinetics, extremely slow changes of lattice parameters were shown in both cells, which suggested the slower delithiation rate and the higher residual lithium content. Therefore, the capacity contributed from $H_{slu}$ might be low until the complete activation of the cathode.

## Discussion

**Two-phase-like structure induced by heterogeneous delithiation**

Since changes of lattice parameter in essential originated from the strain induced by delithiation, we then focused on the total lithium content, i.e., sum of lithium at both 3*a* and 3*b* sites, in two phases and the variation during cycling. In the two-phase-like region of 811|G, $H_{act}$ had a lower lithium content, in which lithium kept migration during further charging, while $H_{slu}$ maintained a high lithium content (> 0.7) until SOC > 80%. Only at high voltage state, where serious *c*-axis collapse occurred in $H_{act}$, as displayed in Fig. 2c and Fig. 3a, $H_{slu}$ was likely to contribute slightly more capacity. But it still held a lithium content of ~0.5 at the end of charge. During discharging, lithium content of $H_{slu}$ approached to that of $H_{act}$, implying the merge of two phases. With the lithiation process, reflections of $H_{act}$, as well as the lithium content and lattice parameters, overlapped with those of $H_{slu}$, exhibiting a solid-solution behavior during the rest of discharging.

In 811|GSO, evolution during two-phase-like region differed from that in 811|G. First, in the two-phase-like region, $H_{slu}$ also showed a lower lithium content than the average at first. However, with the SOC increase, lithium content between $H_{act}$ and $H_{slu}$ approached to each other and soon $H_{slu}$ vanished. The end of two-phase-like region was accompanied by the disappearance of lithium heterogeneity, implying that the two-phase-like phenomenon in essential is the heterogeneous lithium distribution in cathodes.

Additionally, the average lithium content in 811|G varied almost linearly during cycling, except at the start and the end of two-phase-like region, implying a constant delithiation rate during cycling. But a two-segment-like delithiation was observed in 811|GSO. Before the stage of 60% SOC, the cathode presented a slower delithiation rate compared to that in 811|G, while it accelerated when it was over 60% SOC. During delithiation stage, it also slowed down after SOC was reduced to ~60%. This performance could be explained by the reduced dominance of silicon-based particles to the total capacity at high SOC stage.[34] Delithiation process was sped up at the later stage with the increased contribution of graphite. From the results, it was noted that lithium heterogeneity was less-pronounced in 811|GSO under a lower delithiation rate, which revealed that the reaction rate was accounted for the heterogeneous behavior. Under slow reaction process, less $Li^+$ were required to migrate, allowing the sufficient $Li^+$ diffusion in cathode particles to reach the equilibrium state with uniform lithium distribution and eliminate the phase heterogeneity. This mechanism is the same as that of the reduced gradient structure found in the graphite part of GSO. Therefore, a slower reaction process is more favorable for the homogenization of electrode structures.

It should also be noted that cathode in 811|G still maintained a two-phase-like state at the ~100% SOC. The $H_{act}$ phase had contracted *a* and *c* of 2.813(6) Å and 13.88(3) Å as well as the low lithium content of 0.27(4), while lattice parameters (2.861(6) Å for *a*, and 14.24(3) Å for *c*) of $H_{slu}$ phase were close to those before the observation of two-phase-like phenomenon. But in 811|GSO, cathode had evolved as a solid-solution at that voltage, with *a* of 2.812(6) Å, *c* of 14.30(3) Å, and lithium content of 0.36(4). The cathode *c*-axis of 811|GSO at the end of charge was larger than that of $H_{act}$ in 811|G. So was the lithium content. The less degree of *c*-axis collapse in 811|GSO at 100% SOC illustrated that the overall structural evolution progress fell behind that of $H_{act}$ in 811|G, which corresponding to the peak shift in dQ dV$^{-1}$ curves shown in Fig. S4b, ESI†. It seemed that high-delithiated $H_{act}$ could produce more $Li^+$ deintercalation during discharging. However, the remaining sluggish phase at the high voltage state of 811|G restricted the total capacity. The overall delithiation amount in 811|G, reflecting by the discharge curves shown in Fig. S4a, ESI†, was lower than that in 811|GSO. Over-delithiation

of partial cathode had not achieved a higher initial capacity, but apparent lithium heterogeneity that required more cycles to reach the equilibrium. Deduced from the cycling performance in Fig. S3b, ESI†, and the evolution behavior, phase heterogeneity in 811|G could last for several cycles, with the evolution gradually approaching to the homogeneous case. Although the capacity of 811|G got normal after more cycles, the damage of over-delithiation to some cathode particles could result in potential asynchronous aging of electrodes, threatening the lifespan of the cells. In conclusion, to maintain a longer battery life, it is necessary to mitigate the heterogeneous phase evolution of electrodes during cycling, and finally to achieve an equilibrium delithiation state among particles.

**Spatial phase distribution of cathodes at the particle scale**

Now it is clear that the heterogeneous structural evolution of NCM811 in a full cell within the 1$^{st}$ cycle is essentially the delithiation heterogeneity based on the refinement results. However, the structural information from *operando* neutron diffraction is global, which lack the local structure state of electrodes at a certain position of cells. Spatial distribution of delithiated layered phases inside the cells are still unknown. Why did some cathode particles have priority to delithiate over others? How did the kinetic sluggish phase gradually evolve to the active part? And the last is, what strategies could be done to effectively reduce the heterogeneous delithiation process? To explore the problems, *ex situ* 4D-STEM characterization was applied to investigate the phase distribution of NCM811 cathodes at the level of primary particles.

Fig. 4a exhibits the virtual annular bright field (ABF) image of the NCM811 sample from a charged 811|G cell (at 4.3 V, corresponding to the ~100% SOC in the *operando* neutron experiment). Size of NCM811 primary particles is within the range of 200 nm to 1 μm. The corresponding 4D-STEM phase map is shown in Fig. 4b. For comparation, virtual ABF image and phase map for the pristine NCM811 particles from an uncharged cell were also collected, as shown in Fig. S9, ESI†. Among the undelithiated NCM811 particles, the rhombohedral H$_{slu}$ phase occupies most of the analyzing area, and tiny spinel-like phase is identified in minor regions. Combined with the ABF image, most spinel-like structure is located on or near the surface, which might be accounted for the corrosion from electrolyte, or lithium loss during synthesis/storage. Anyway, it is not easy to prepare NCM811 cathode with purely perfect layered structure. Little amount of the spinel-like phase is neglected in the static-state *operando* neutron patterns. However, for the charged NCM811 particles from NCM811|G, except a trace of surface region, a majority of spinel-like phase is observed in a few particles (Fig. 4b). It is also found that the spinel-like region was adjacent or entangled with the highly-delithiated H$_{act}$ domain, which could be deduced that the spinel-like structure in the particle is the degeneration

product of unstable over-delithiated layered structure. The orientation relationship between spinel-like and the layered phase is determined as: $\langle 111 \rangle_s \parallel \langle 001 \rangle_l$, $\langle \bar{1}10 \rangle_s \parallel \langle 100 \rangle_l$, and $\langle 11\bar{2} \rangle_s \parallel \langle \bar{1}\bar{2}0 \rangle_l$.

As for the $H_{act}$ and $H_{slu}$ phases, both inter-particle and intra-particle phase heterogeneity are observed from Fig. 4b. In the detection region, 20 typical primary particles are picked up, with their size and phase distribution described qualitatively in Table S4, ESI†. From the results, SOC of a single cathode particle could be reflected by the main phase with the largest region. For example, in particles with marked number 5, 10, 14, 15, and 18, the dominated $H_{slu}$ region represented a low charge state/delithiation degree. In particles No. 8 and No. 13, the large-area spinel-like phase region reflected the serious deterioration of layered structure, corresponding to the overcharged state. Others display a high delithiation level with large-distributed $H_{act}$ domains. The difference in delithiation ability among particles might originated from the surface state,[45] the conductive network construction,[48] the electrolyte infiltration,[49] etc. The obvious inter-particle phase heterogeneity is clearly revealed by the 4D-STEM method.

For the intra-particle heterogeneous phase structure, contradicted to the past recognition that delithiation in a single particle is from the surface to the bulk, phase mapping for particles dominated by the $H_{act}$ phase (No. 1, 2, 3, 12, 17 and 19) illustrate that significant content of $H_{slu}$ or spinel-like phases are distributed on the surfaces. Less $H_{slu}$ or spinel-like phase domains are dispersed inside these particles, as shown in Fig. 4a-b. Take particle No. 17 as an example, the $H_{slu}$ region occupies most area of the surface. Thickness of the kinetic inactive $H_{slu}$ shell is within the range of 20 to 50 nm. Compared with the particle size of several hundred nanometers, although area fraction of the $H_{slu}$ domain to the particle is not very large, any delithiation heterogeneity might leave potential hazards to further cell performance. Therefore, hidden mechanisms that induce the $H_{slu}$ phase distribution bias in a particle requires investigations.

**Excluded factors resulting in delithiation heterogeneity**

Several hypotheses are put forward and tested. One possibility is the size effect, i.e., lithium heterogeneity is more pronounced in larger particles due to the longer Li$^+$ diffusion path.[42,44] However, few evidences support that in this work. According to results in Fig. 4a-b and Table S4, ESI†, whatever small (No. 7), medium (No. 3, 17), or large (No. 2, 12) particles, $H_{slu}$ domains colored by orange are clearly shown on surface regions with thickness of dozen nanometers. This finding suggests poor relevance between size and phase heterogeneity. The random orientation of cathode particles in Fig. S10, ESI†, excludes the possibility of particle orientation to delithiation preference. The composition-induced delithiation heterogeneity is also excluded since the large-area EDS mappings of delithiated sample do not exhibit apparent element segregation on particle surfaces in Fig. S11, ESI†. The overall uniform element distribution in the charged sample could not lead to the delithiation bias in a single particle.

**Delithiation influenced by the distribution of Li/Ni disordering**

Herein, we propose the dominant mechanism that lithium migration from cathode particles is led by the evolution of defects, particularly the known Li/Ni disordering for the Ni-rich family. Distribution of defects could result in the delithiation ability difference among each position of a particle. In the bulk, distribution of Li/Ni disordering is dispersed according to our observation. In Fig. 4c, the atomic resolution high-angle annular dark field (HAADF) image clearly exhibited the typical layered structure of an uncharged NCM811 sample at [100] zone axis. Layers of bright spots, representing TM ions, repeat every 3 stacking layers along *c*-axis. However, in addition to the common features, slight bright dots are visible between neighboring TM layers, where lithium ions located. Z contrast of light elements like Li or O is invisible in HAADF STEM images. The observed weak contrast should be attributed to the disordered Ni at Li sites. Moreover, intensity of lithium sites varies significantly with the location, indicating the non-uniform Li/Ni disordering distribution. Three selected line scans are plotted in Fig. 4d, in which TM ions and lithium ions arranged alternatively. Comparing the intensity at lithium sites with neighboring TM sites, significant fluctuation was observed along scan paths. Some lithium sites which should have low intensity display higher brightness (marked with an orange upper triangle), corresponding to a local high Li/Ni disordering content, while some have a lower intensity than the averaged (marked with a purple lower triangle), representing a local ideal layered structure. Images of layered NCM811 illustrated that the bulk Li/Ni disordering distribution could be rather dispersed but heterogeneous, i.e., under a domain with extremely small size, the Li/Ni disordering content might diverge markedly from the average. Besides, complex phase structures with the rock-salt-like, disordered layered and ordered layered areas are found at different depths from the surface, indicating the concentration of Li/Ni disordering near the particle surface, as shown in Fig. 4e. Size of rock-salt-like shell is smaller than that of the pixel size in 4D-STEM mappings, which is hard to be identified in Fig. 4b.

The dispersed distribution of Li/Ni disordering in the bulk and the concentration near the surface are consistent with the distribution of $H_{slu}$ phase in highly-delithiated particles, which reveals the relationship between heterogeneous phase/lithium distribution and defect distribution. During charging, intra-particle phase/lithium heterogeneity is guided by the Li/Ni disordering distribution. Domains with less Li/Ni disordering content tend to delithiate preferentially, due to the slower lithium diffusion in the layered structure with more Li/Ni disordering.[50] Anti-site Ni ions could influence the two-dimensional lithium migration path in the lattice by introducing stronger coulombic repulsion from Ni ions and local distortion, leading to the hindered lithium diffusion and slower kinetics in disordered structures. With further delithiation, difference in lithium diffusion between ordered and disordered domains finally manifests as the phase heterogeneity in a charged cathode.

Evolution of Li/Ni disordering content for $H_{act}$ and $H_{slu}$ phases in Fig. S12, ESI†, during cycling clearly verifies the anti-site-driven two-phase-like process. The proposed intra-particle delithiation process is sketched in Fig. 4f. At the beginning of charge, reaction occurred on individual sites of the surface. The initial delithiation kinetics was slow since migration in the shallow layer of particles could be affected by surface defects. With the ongoing lithium deintercalation, diffusion gradually deepened into the bulk, facilitating the delithiation in ordered layered domains with a relative faster speed. Then the lithium-poor $H_{act}$ phase with larger lattice parameters variation nucleated in ordered layered domains near the surface. The composition gradient between $H_{act}$ nucleus and neighboring lithium-rich domains triggered the lithium migration from these domains and the growth of $H_{act}$ phase. Next, sufficiently large difference in lattice parameter and lithium content between $H_{act}$ and the rest kinetic sluggish $H_{slu}$ part could be observed in Fig. 3. When $H_{act}$ was distinguished by neutron diffraction, the Li/Ni disordering of that was lower than the average value, as shown in Fig. S12, ESI†. In further delithiation stage, the average Li/Ni disordering amount first increased then decreased to a low value in both cells, which is due to the competition between enhanced lithium vacancy content and reduced $Ni^{2+}$ amount as illustrated in our previous work.[35] Simultaneously, $H_{act}$ reflections became stronger in Fig. S7, ESI†, revealing the continuous growth of $H_{act}$ phase. During the expansion, Li/Ni disordering of $H_{act}$ approached to the average value in both cells, but did not exceed it. This result confirms that $H_{act}$ always contained relatively lower Li/Ni disordering during cycling. Based on this, it is reasonable to consider that lithium ions in domains with less Li/Ni disordering are more favorable to be extracted first during delithiation. Ordered layered domains in uncharged particle will gradually evolve to $H_{act}$ phase. And the particle surface and some highly disordered domains, which is negative for lithium migration, maintain the original structure and delithiate last, or leave as the $H_{slu}$ phase shown in Fig. 4f.

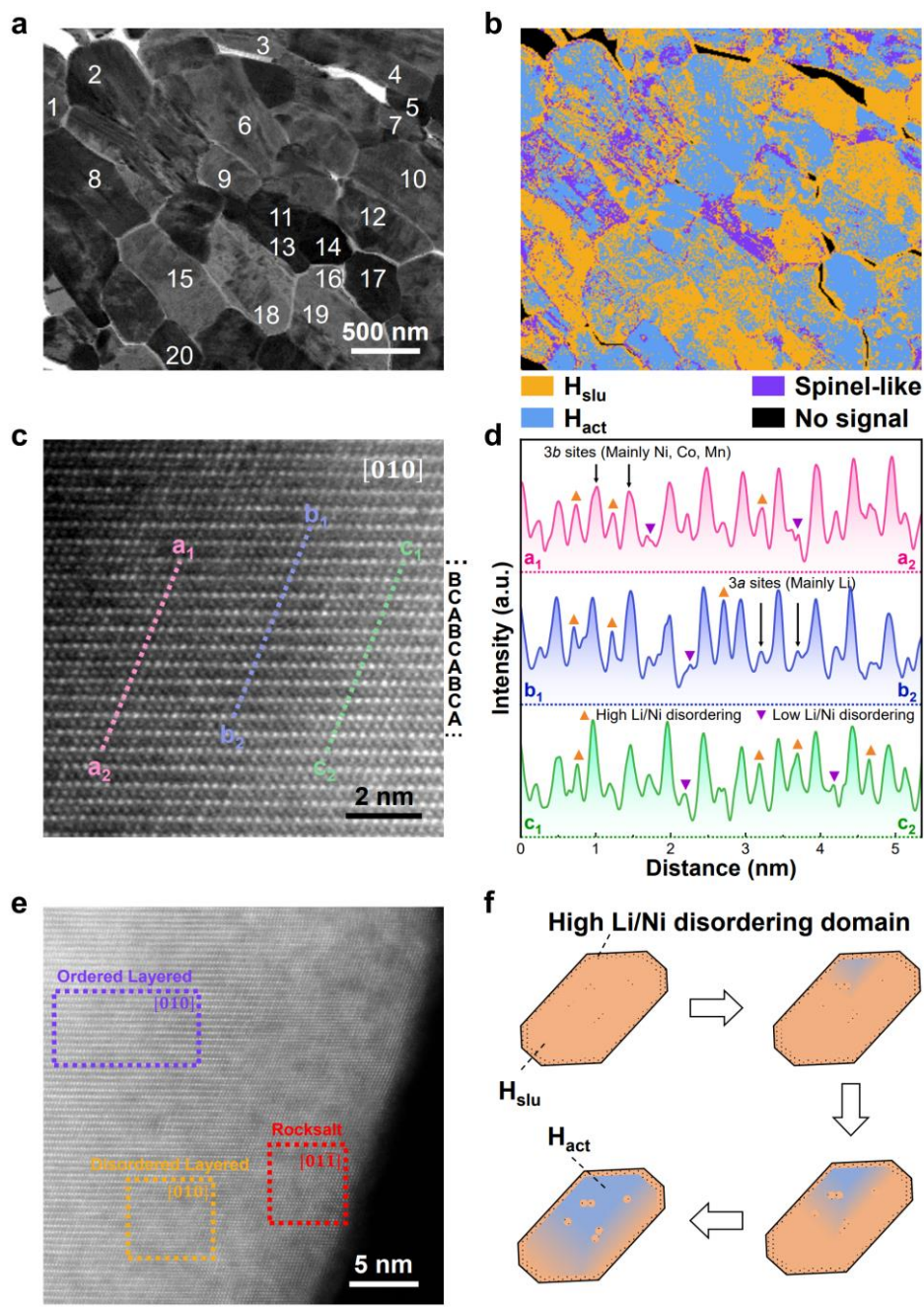

**Fig. 4** Phase heterogeneity in delithiated NCM811 cathode at the primary particle scale driven by Li/Ni disordering. (a) Virtual ABF image of NCM811 primary particles from 811|G after the 1st charge to 4.3 V. (b) Phase map revealing the distribution of $H_{slu}$, $H_{act}$, and spinel-like phase in cathode particles shown in (a). Pixel size: 9.9 nm × 9.9 nm. (c) Atomic resolution HAADF image of pristine NCM811 bulk structure in [010] zone axis, in which brighter contrast of an atomic column represents higher occupancy of TM ions. (d) Selected line profiles of atomic contrast from (c). (e) Atomic resolution HAADF image of a charged NCM811 particle near the surface. Representative phase structures are framed by dash lines. (f) Proposed activation process influenced by the distribution of Li/Ni disordering.

**Delithiation influenced by the formation of microcracks**

Besides the Li/Ni disordering, another defect, known as microcracks,[2,51,52] also influence the spatial phase distribution. Fig. 5a-b are the 4D-STEM phase mapping of a single particle, with large area of $H_{act}$ penetrating the middle of the particle. Minor dispersed $H_{act}$ pixels are also found near the surface. Moreover, strip-shape zones transversing the particle are discovered in both Fig. 5a and Fig. 5b. Strain mappings of the particle in Fig. 5c-d clearly reveal that the strip-shaped zones are intragranular microcracks with significant strain concentration. Strong normal tensile strain in x-direction ($\varepsilon_{xx}$, x ∥ c-axis) is observed along microcracks, while normal strain in y-direction ($\varepsilon_{yy}$, y ⊥ c-axis) is much lower. No obvious shear strain ($\varepsilon_{xy}$) is found in the particle, as displayed in Fig. S13, ESI†. Similar phenomenon does not appear in the uncharged particle (Fig. S14, ESI†), suggesting the strain accumulation during delithiation. The tensile strain will break the ordered layered structure, especially the weakened interlayer connection between lithium and neighboring oxygen layers at high SOC. Over-expanded lithium layers finally become the nucleation sites of intergranular microcracks, as exhibited in Fig. 5e. Near microcracks, domains with high Li/Ni disordering content are observed (marked with a pink rectangular). Since Li/Ni disordering could induce lattice mismatch or local distortion by inhibiting the *c*-axis expansion during delithiation, strain is more favorable to concentrate near/in disordered domains. Then the condition for microcrack appearance is satisfied.[53]

    Combing the phase and strain/microcracks distribution in Fig. 5b-c, it is noticed that some intergranular microcracks penetrate the particle as well as large-area $H_{act}$ phase inside the particle. Although the $H_{act}$ region is connected to the particle surface, area of that located in the bulk is larger than that near the surface. It could be inferred the facilitation of microcracks to the phase evolution. That is, transgranular microcracks would promote the expansion of delithiated $H_{act}$ phase by extra interfaces for electrolyte contact and lithium transportation. As proposed in Fig. 5f (neglect effects from other defects), once a transgranular microcrack form, it could lead the nucleated $H_{act}$ phase to grow near the newly generated interface preferentially, then extending to other positions. It should be acknowledged that in practical cell systems, especially the large-size energy storage devices, internal defects as well as other external factors[42,44-46,48,49] would influence the structural evolution simultaneously, resulting in a more complicated multi-scale phase/lithium distribution. This work focuses on the scale of primary particles, and figures out that the intraparticle phase heterogeneity is mainly induced by defects including Li/Ni disordering and microcracks.

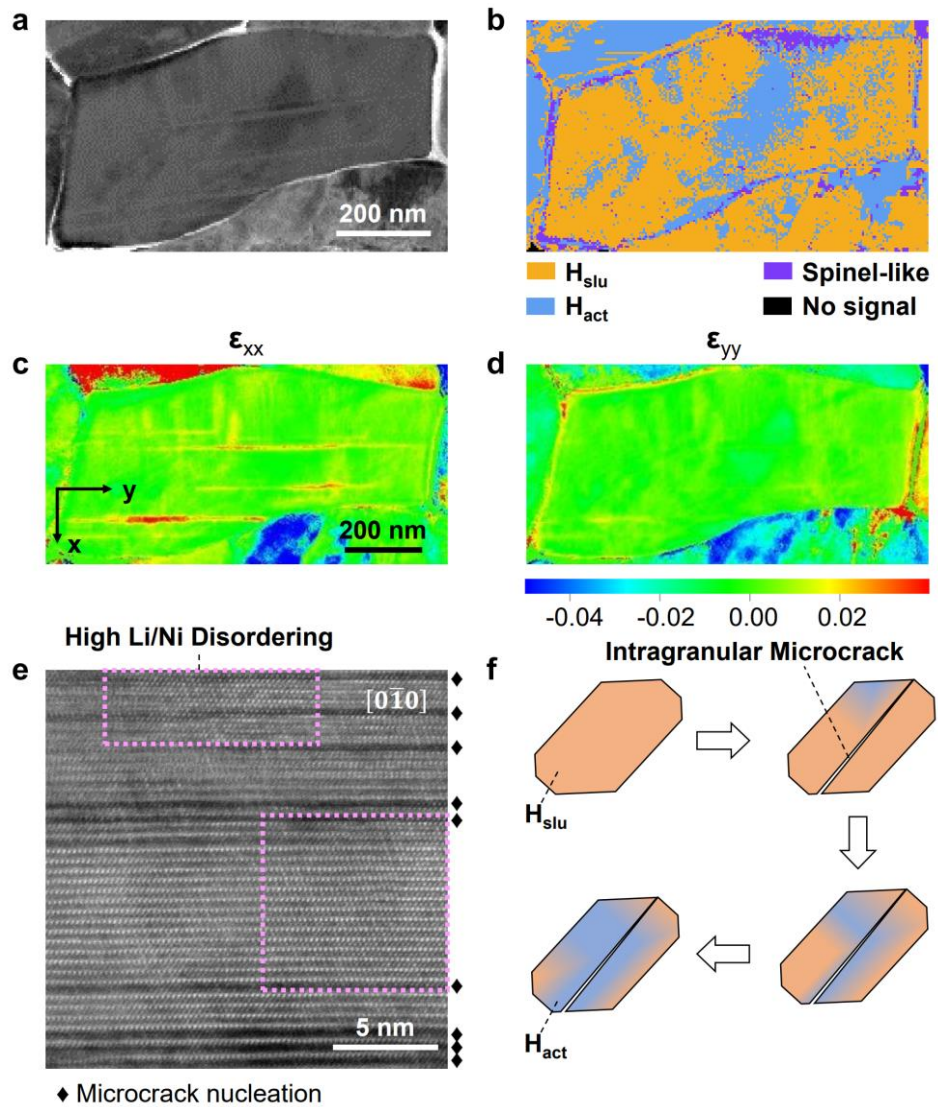

**Fig. 5** Phase heterogeneity in a NCM811 primary particle driven by microcracks. (a) Virtual ABF image of a charged NCM811 particle with intragranular microcracks from 811|G. (b) Phase map of the particle shown in (a). Pixel size: 3.5 nm × 3.5 nm. Distribution of the normal strain in (c) x-direction, $\varepsilon_{xx}$, and (d) y-direction, $\varepsilon_{yy}$, in the selected particle shown in (a). (e) Atomic resolution HAADF image of charged NCM811 bulk structure with nucleation of intragranular microcracks marked by "♦", and domains framed by dash lines with high Li/Ni disordering content. (f) Proposed activation process influenced by the intragranular microcrack formation (ignore other defects in the particle).

**Effects of slow kinetics on the defect-driven delithiation process**

It is also noted that degree of heterogeneity in two cells during the first cycle is different. NCM811 cathode with graphite anode experienced a longer and more serious heterogeneous evolution process compared to that with GSO anode, which should be ascribed to the slow kinetics of silicon-based particles. Electrochemical properties of anode could influence the lithium migration in the cell system. The fast lithium intercalation of graphite requires fast delithiation from cathode. But for layered cathode, lithium ions located in domains with less neighboring Li/Ni disordering or close to microcracks could support the demand. As a result, after the initial lithium deintercalation from these domains, further delithiation from the kinetic sluggish part ($H_{slu}$) with more Li/Ni disordering or far from the surface/interface cannot satisfy the fast reaction rate. Then the kinetic active part ($H_{act}$) is forced to contribute more lithium, leading to the non-equilibrium composition/structure of cathode. Besides, the heterogeneous phase distribution and the gradually increased difference in lattice parameters will aggravate the strain accumulation and microcracks formation in highly delithiated particles. Microcracks then promote the lithium extraction. This feedback causes the phenomenon that lithium-poor phase becomes more lithium-poor, like the Matthew effect in sociology. But in the cell with slow kinetics, the $H_{slu}$ phase could catch up the reaction process, and reach the equilibrium structural state. No obvious lithium content is found between domains with high and low Li/Ni disordering content. The relative uniform lithium distribution provides close condition for microcrack growth, resulting in the same crack density for each primary particle. In summary, a slower electrochemical process will significantly suppress the heterogeneous lithium migration in electrodes, promoting the uniform phase evolution and stable electrochemical process.

In practical situation, it is necessary to consider the kinetic properties of electrodes to guarantee the longevity, stability, and safety of cells. Therefore, necessary strategies should be taken to reduce heterogeneity behaviors in cells during cycling to avoid the influence to cell capacity and cycle life. For example, adjusting the anode composition to control the cell kinetics could be one simple and useful method as revealed in this work. However, in case of fast charge, slow delithiation process strongly restricted the application of silicon-based anodes. Further modifications,[54] including electrode nano-crystallization, conductive network engineering, electrolyte design, etc., are all required to systematically improve the rate performance. Also, the very carefully synthesis of electrode materials, especially the Ni-rich cathode, could effectively reduce the defect content and enhance the cell performance. In any case, realization of high-performance LIBs is quite a challenge that needs to consider multiple aspects.

## Conclusions

In summary, this work gets insight into the phase heterogeneity phenomenon in electrode materials during the initial cycling from both the global view via *operando* neutron diffraction and the local view via *ex situ* 4D-STEM characterizations. The neutron probe clearly reveals the bimodal distribution of cathode structure and its evolution with the time. Extracted structural information indicates that the essence of phase heterogeneity is the non-uniform lithium distribution. Phase and strain mappings from 4D-STEM provides the important complementary information about the spatial distribution of phases in cathode primary particles. And lattice defects including Li/Ni disordering, microcracks, etc., are found to be one internal mechanism that drives the heterogeneous lithium deintercalation at the particle scale. Based on the findings above, it is suggested to inhibit the non-intrinsic phase heterogeneity by careful optimization for the cell structure or controls of the kinetic process. Our work not only sheds light on the methodology for structural investigations in complex full cell systems, but also supply critical experience for the further development of LIBs with higher capacity and longer life.

## Author Contributions

Yinguo Xiao conceived and supervised this project. Zhongyuan Huang, Rui Wang, Mihai Chu, Ziwei Chen, and Maolin Yang performed the preparation and characterization of electrodes as well as the electrochemical measurements. Zenan Li and Wang Jun performed the fabrication of full cells. Zhongyuan Huang, Rui Wang, Mihai Chu, Sihao Deng, Lunhua He, and Yinguo Xiao designed and performed the *operando* neutron diffraction experiments. Zhongyuan Huang performed the structural analysis of the *operando* neutron data under the guidance of Yinguo Xiao. Tingting Yang, Lei Jin, and Rafal E. Dunin-Borkowski conducted 4D-STEM characterizations and guided the data analysis. Zhongyuan Huang wrote the manuscript with contributions from all authors. All authors discussed the manuscript and proposed suggestions for revision.

## Conflict of Interest

There are no conflicts to declare.

## Acknowledgements

This work was financially supported by the National Key R&D Program of China (2020YFA0406203), National Natural Science Foundation of China (Nos. 52072008, and U2032167), Guangdong Basic and Applied Basic Research Foundation (Nos. 2022B1515120070, and 2022A1515110816), Shenzhen Fundamental Research Program (No. GXWD20201231165807007-20200807125314001), and the Large Scientific Facility Open Subject of Songshan Lake, Dongguan, Guangdong (No. KFKT2022A04). The Major Science and Technology Infrastructure Project of Material Genome Big-science Facilities Platform supported by Municipal Development and Reform Commission of Shenzhen. The authors appreciate the neutron beamtime at GPPD granted from CSNS, Dongguan, China. The authors appreciate Yaoda Wu for technical assistance during *operando* neutron diffraction experiments.

Electronic Supplementary Information (ESI) for

# Insights into the defect-driven heterogeneous structural evolution of Ni-rich layered cathode in lithium-ion batteries

Zhongyuan Huang,[a] Ziwei Chen,[a] Maolin Yang,[a] Mihai Chu,[b] Zenan Li,[c] Sihao Deng,[de] Lunhua He,[efg] Lei Jin,[h] Rafal E. Dunin-Borkowski,[h] Rui Wang,*[i] Jun Wang,*[c] Tingting Yang,*[h] and Yinguo Xiao*[a]

[a] School of Advanced Materials, Peking University, Shenzhen Graduate School, Shenzhen 518055, PR China

[b] Department of Chemistry, Materials, and Chemical Engineering "Giulio Natta", Politecnico di Milano, Milano 20133, Italy

[c] School of Innovation and Entrepreneurship, Southern University of Science and Technology, Shenzhen 518055, PR China

[d] Institute of High Energy Physics, Chinese Academy of Sciences, Beijing 100049, PR China

[e] Spallation Neutron Source Science Center, Dongguan 523803, PR China

[f] Beijing National Laboratory for Condensed Matter Physics, Institute of Physics, Chinese Academy of Sciences, Beijing 100190, PR China

[g] Songshan Lake Materials Laboratory, Dongguan 523808, PR China

[h] Ernst Ruska-Centre for Microscopy and Spectroscopy with Electrons, Forschungszentrum Jülich GmbH, Jülich 52428, Germany

[i] Department of Engineering, University of Cambridge, Cambridge CB30FS, UK

* Corresponding Authors: Jun Wang (wangj9@sustech.edu.cn), Rui Wang (rw716@cam.ac.uk), Tingting Yang (t.yang@fz-juelich.de), Yinguo Xiao (y.xiao@pku.edu.cn)

# Experimental

## Electrode preparations

Layered NCM811 cathode powder was prepared via a solid-state reaction of $Ni_{0.8}Co_{0.1}Mn_{0.1}(OH)_2$ precursor and $LiOH·H_2O$. Detail process is reported in the previous article.[1] Next, the powder product was dispersed in *N*-methyl-2-pyrrolidone (NMP) with acetylene black and poly- (vinylidene fluoride) (PVDF) at the ratio of 90:5:5 in mass to make the slurry. The slurry was cast onto aluminum foil and dried at 110°C for 12 h under vacuum to prepare NCM811 cathode.

Silicon-based composite was prepared by a magnesio-mechanochemical reaction of diatomite and Mg powder. After the reaction, Mg-containing byproducts were removed by HCl etching. The obtained product was then coated with carbon. Details could be found in previous works.[2,3] Finally, the material consisted of mixed silicon and silica nanodomains and carbon shells, whose mass ratio was near 1:1:1. Two types of anode powder were prepared. The first is pure graphite, and the latter is graphite-silicon-silica (GSO) powder with about 95% graphite (from Kejing, Shenzhen, China) and about 5% as-prepared carbon-coated silicon-based composite in mass by ball-milling. The anode powder, carbon black, and polyacrylic acid (PAA, $\overline{M_v}$~450,000) were stirred in deionized water for 5 h with the 8:1:1 mass ratio, then cast onto copper foil and dried under the same condition of the cathode to prepare the pure graphite or GSO anodes.

## Full cell assembles and electrochemical measurements

Full cells were assembled in an Argon-filled glovebox, all of which used the NCM811 cathode and Celgard 2400 separators. The electrolyte was 1 M $LiPF_6$ dissolved in solvent with propylene carbonate (PC), fluoroethylene carbonate (FEC), diethyl carbonate (DEC) under 1:1:4 in volume, and 1% vinylene carbonate (VC) in mass. The injection volume was minimized to reduce the neutron incoherent scattering of H from organic solvents, which guaranteed the quality of *operando* neutron patterns. Negative/positive electrode capacity ratio (N/P) of cells was controlled around 1.1. Cells had the electrode area of 62.7 × 50.0 mm$^2$ and the thickness of ~3.63 mm, which contained 16 parallel stacking layers of cathode, anode, and separators to enhance the Bragg scattering signal from electrodes. Two types of cells with different anodes were assembled. One type used the pure graphite anode, named as 811|G. Another chose the GSO anode, named as 811|GSO. A batch of 5 basically identical cells were assembled for each type, which would be used for electrochemical tests (rate and cycling performance) and further research including *operando* neutron experiment, 4D-STEM, etc. Two cells of each type were used for galvanostatic discharge-charge tests between 2.7 and 4.3 V on a NEWARE battery test system. One was tested under different rates (0.1C to 1C, 5 cycles for each). Another was charged and discharged once at 0.1C and then cycled at 0.5C.

*Operando* **neutron diffraction experiments**

*Operando* neutron diffraction experiments were carried on the General Purpose Powder Diffractometer (GPPD)[4] at China Spallation Neutron Source (CSNS), Dongguan, Guangdong, China. GPPD is a versatile time-of-flight (TOF) diffractometer with 3 sets of detector banks, of which centers are located at $2\theta = 30°$, $90°$ and $150°$ respectively. It supports research needs of various materials, as well as special environments for *operando* structure studies of cycling cells. Obviously, counting rate is the most critical requirement of an *operando* experiment since it decides the time resolution of a pattern. The gradual upgrade of CSNS facility and efforts from staff permit high enough flux of GPPD to collect the neutron data within acceptable time. Moreover, to maximize the signal collection, beam size was set as 40 mm in height and 20 mm in width.

Before the experiments, 811|G and 811|GSO cells were hanged on the scattering cavity by a sample holder with insulated clamp fixture. Distance from the upper end of the holder to the cell center was about 80 mm. Wires were wound around the holder, and covered with neutron shielding materials to avoid unnecessary signal. One end of the wire was connected to the cell, while another end was connected to a LANHE CT2001A Battery Testing System. After the installation of equipment and the cell, a series of tests was carried out, including checking the cell state, testing external circuit, collecting a pattern for the cell at the static state, etc., to ensure the smooth progress of following experiments.

Before charging the first cell, a 1-hour trial testing was carried out for data quality evaluation. Based on the test result for the uncharged cell, 20 min was decided for one pattern collection. A time gap of ~3 min was left between neighboring patterns for data saving and starting the next. So, the total data collecting time for one pattern is ~23 min. Cells were charged from OCV to 4.3 V under 0.2C, then discharged to 2.7 V (versus corresponding anodes) with the same rate. Charge time of 811|G and 811|GSO during experiments are 5.04 and 5.15 h, while discharge time are 3.85 and 3.99 h, respectively.

**Data processing and Rietveld refinements**

SOC of each *operando* neutron pattern is determined as follows: First, the static state before charge was defined as the 0% SOC, and the time when the 1$^{st}$ charge was finished was defined as the 100% SOC. Then the SOC of a pattern (suppose the $i^{th}$ one, $i = 1, 2, 3…$) during charging could be determined by:

$$SOC_{i,cha.} = \frac{q_{i,cha.}}{q_{cha.}} = \frac{I_{cha.}t_{i,cha.}}{I_{cha.}t_{cha.}} = \frac{t_{i,cha.}}{t_{cha.}}. \qquad (1)$$

In the definition, $q_{i, cha.}$ is the accumulated charge capacity at the mid-time point when the $i^{th}$ pattern was collecting. $q_{cha.}$ is the total charge capacity. $I_{cha.}$ is the charge current. $t_{i, cha.}$ and $t_{cha.}$ are the mid-time point of the $i^{th}$ pattern collection and the total charge time, respectively. Similarly, SOC of patterns during discharging could also be defined by:

$$SOC_{i,dis.} = 1 - \frac{q_{i,dis.}}{q_{cha.}} = 1 - \frac{I_{dis.} t_{i,dis.}}{I_{cha.} t_{cha.}} = 1 - \frac{t_{i,dis.}}{t_{cha.}}. \tag{2}$$

Since the discharge current $I_{dis.}$ is equal to the charge current, which is 0.2C, the capacity ratio is also converted to the time ratio. It is worthing noting that the total charge capacity is still used as the reference of 100% SOC in this work, rather than the total discharge capacity. The reason is that the crystal structure of electrodes at the end of 1$^{st}$ discharge is usually different from those at the uncharged state due to a degree of irreversibility. If the total discharge capacity were the denominator, value of SOC for the end state would be almost equal to that for the state before charge (0%), leading to a misunderstanding that the final structure might be almost the same as the initial. To avoid that, the 1$^{st}$ charge capacity is used as the standard to calibrate the SOC for neutron data. Additionally, the last data for each cell, which was acquired at almost the end of discharge and a period of quiescent state, was not be assigned the SOC value because it was under neither charge nor discharge.

$d$-spacing in Å of each neutron data point was determined by converting TOF in μs by:

$$TOF = Z_0 + Dtt_1 d + Dtt_2 d^2. \tag{3}$$

In this formula, $Z_0$ is the constant shift. $Dtt_1$ is the diffractometer constant calibrated by the standard sample. $Dtt_2$ is the correction coefficient for sample displacement and absorption-induced shifts. Since cells and holders were not moved during *operando* experiments, $Z_0$ and $Dtt_2$ for each cell were determined by refining the patterns collected under the uncharged state. The three parameters were fixed when doing structure refinement for other *operando* patterns.

Rietveld refinements were carried out for the neutron data between 0.75 and 2.6 Å through the FullProf Suite.[5] A set of data points located at the position where no Bragg reflection was observed were selected and connected linearly to fit the background. Background points were allowed for optimization during refinements. A Pseudo-Voigt function convoluted with a back-to-back exponential was adopted to describe the peak profile. Peak profile parameters from the instrument resolution file were used as the initial value, which were further optimized. Both the isotropic size and strain parameters were adjusted to match the FWHM of reflections in *operando* patterns during refinements. The Lorentz and polarization factors were accounted by multiplying $d^4$ and $\sin \theta$. Based on the comparison of the refined *operando* data to that from previous *operando* researches[6] or powder samples,[7] refinement results were not noticeably influenced, suggesting that the model could well match the experiment data.

Steps of refinements were shown below: First, scale factors, lattice parameters, background parameters, profile parameters, etc. were refined and optimized to appropriate value. Next, they were fixed and structure parameters, like *z*-coordinate of O at 6*c* site, occupancy of Li at 3*a* site, Li/Ni disordering amount of cathode, etc., were allowed to vary. Correlation of structural parameters, which was reduced by strict constrains including fixing atomic displacement parameters, reducing the variation of refined parameters, and restricting the variation ranges. Last, the obtained results were used as the initial value for the refinement of the next. The same steps would repeat for each data. Parameters of current collectors, which did not evolve during cycling, were not refined from the 2$^{nd}$ pattern. Refined structures of crystallized phases in electrodes were visualized by the VESTA program.[8] Contour plots for the two sets of full cell *operando* neutron diffraction data after background subtractions were realized by MATLAB.

**4D-STEM and other electron microscopy characterizations**

A ZEISS Supra 55 field-emission scanning electron microscope (SEM) equipped with energy dispersive spectrometer (EDS) was used to observe the morphologies and analyze the elements of NCM811 cathode, pure graphite anode, and GSO anode.

One 811|G cell was charged to 4.3 V and disassembled in the Argon-filled glove box. NCM811 powder from two cells was then carefully scaped from the Al foil and washed by DEC to reduce residual salts. After drying, the obtained powder was sealed and used for electron microscopy characterizations. The atomic HAADF imaging was conducted on a FEl Titan G2 ChemiSTEM 80-200 transmission electron microscope operated at 200 kV. The microscope is equipped with a high-brightness field emission gun and a probe aberration corrector for aberration-corrected STEM. Tomography EDS was conducted on a TFS Spectra 300 (S)TEM operated at 300 kV, and the tilting angle was between -55° to 55°. The microscope is equipped with an ultra-high brightness field emission gun, a probe aberration corrector for aberration-corrected STEM and an image aberration corrector for aberration-corrected TEM. 4D-STEM experiment was conducted on Tescan Tensor microscope. which is the world first precession assisted 4D-STEM microscope and can realize near real-time analysis and processing of 4D-STEM data. The 4D-STEM was performed at a convergence semi-angle of 1.5 mrad, beam current of 200 pA, and probe size of 1.5 nm. The diffraction size is 124.5 mrad. The precession angel is 0.8°. The electron probe was raster-scanned across the selection area using a step size of 3.5 nm and a diffraction pattern recorded at each probe position with a high-performance, hybrid-pixel, direct electron diffraction camera (Dectris Quadro). The camera has 512 × 512

physical pixels, used for the strain measurements. Additionally, 4-fold (128 × 128 pixels) pixel binning is utilized for orientation and phase mapping.

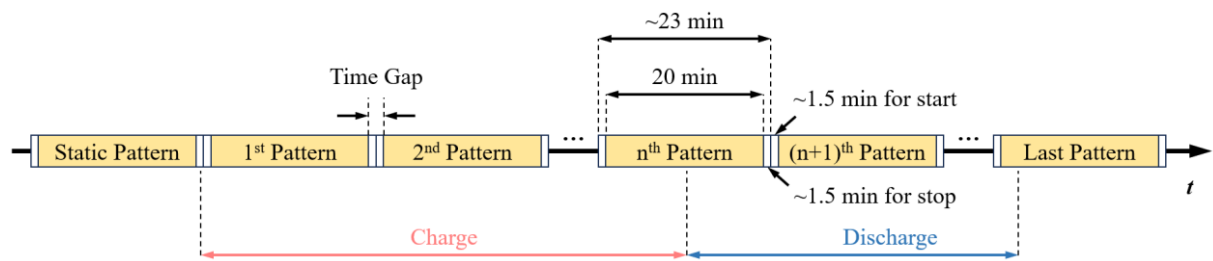

**Fig. S1** Schematic of data collection plan and the corresponding electrochemical state of cells during the *operando* neutron experiments.

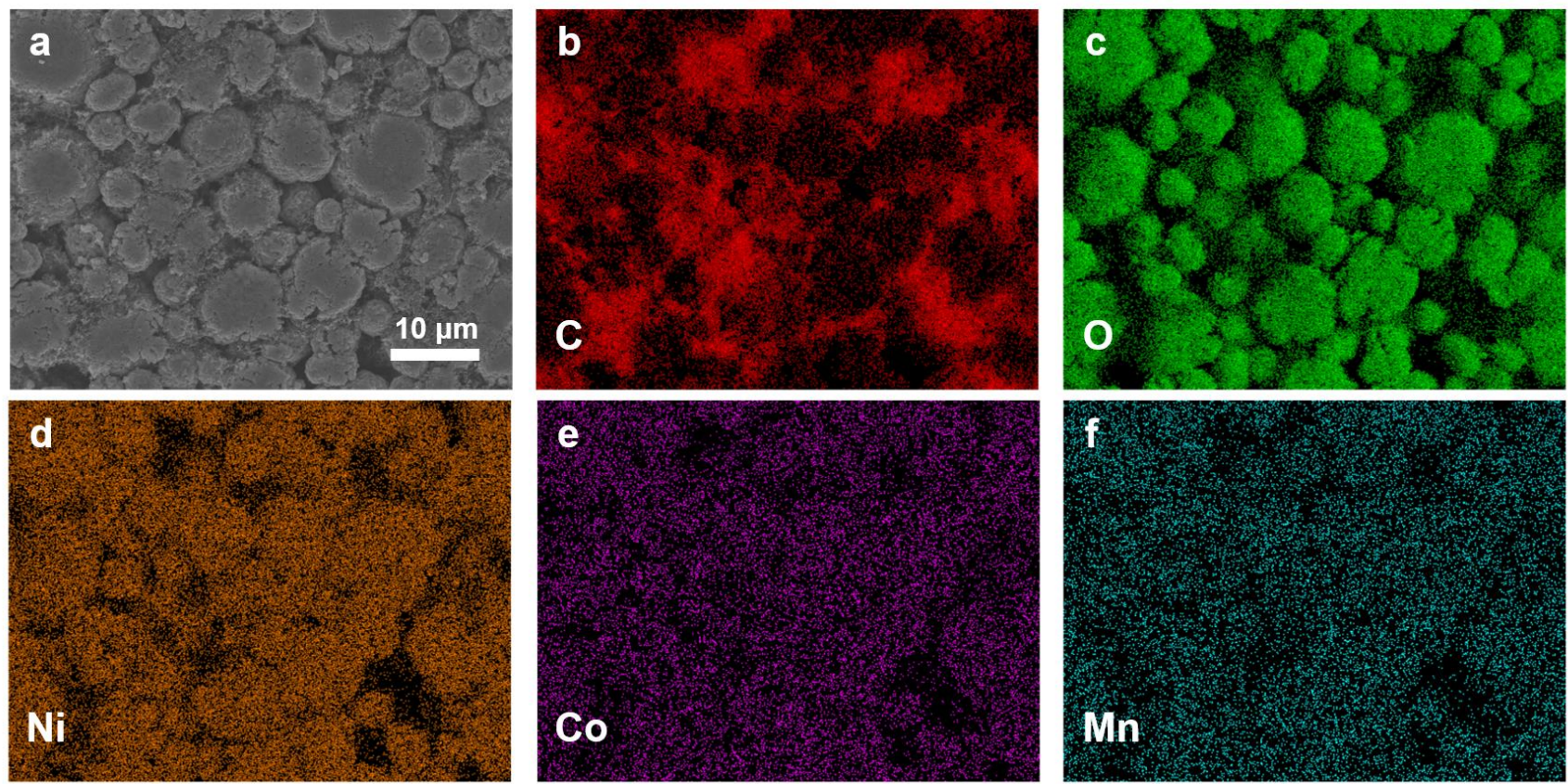

**Fig. S2** (a) SEM morphology image of NCM811 cathode and the corresponding EDS mappings of (b) carbon, (c) oxygen, (d) nickel, (e) cobalt, and (f) manganese.

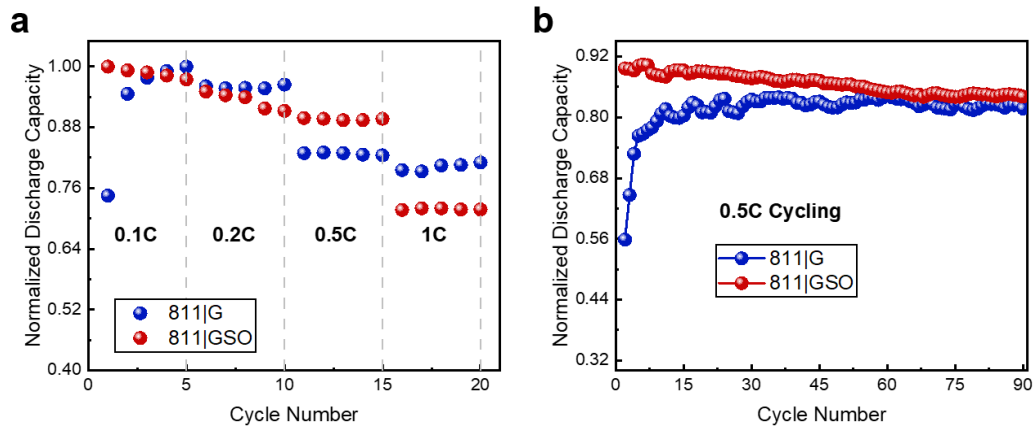

**Fig. S3** (a) Rate performance of 811|G and 811|GSO from 0.1C to 1C. (b) Cycling performance of the full cells at 0.5C. Discharge capacity was normalized through dividing by the maximum capacity under 0.1C. Voltage is within the range of 2.7~4.3 V.

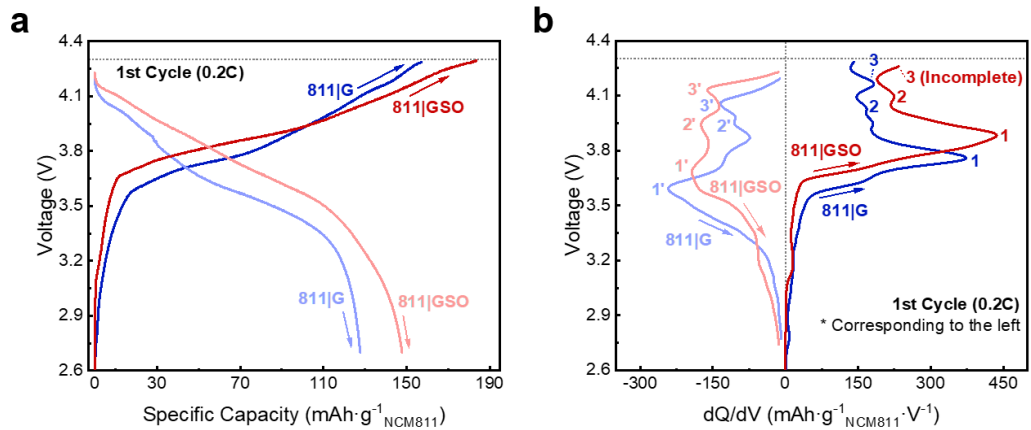

**Fig. S4** (a) The 1st 0.2C cycling charge-discharge curve of 811|G and 811|GSO cells between 2.7 and 4.3 V during the *operando* neutron diffraction experiments. (b) Corresponding dQ·dV$^{-1}$ curves of 811|G and 811|GSO cells. Peaks with assigned numbers represent the so-called "phase transitions" reported in previous studies. (1: H1 → M; 2: M → H2; 3: H2 → H3; 1', 2', and 3' are the inverse processes.) Some peaks split due to the inhomogeneous electrochemical delithiation/lithiation during the 1st cycling.

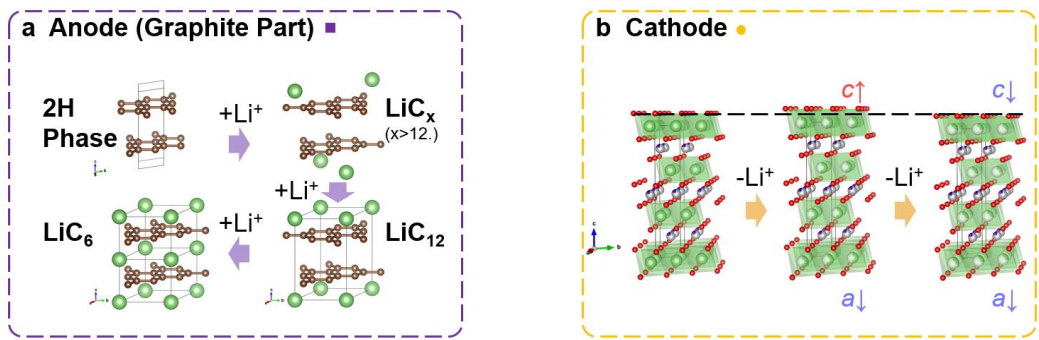

**Fig. S5** (a) Schematic of graphite structure transition during lithiation. (b) Schematic of NCM811 structural evolution during delithiation.

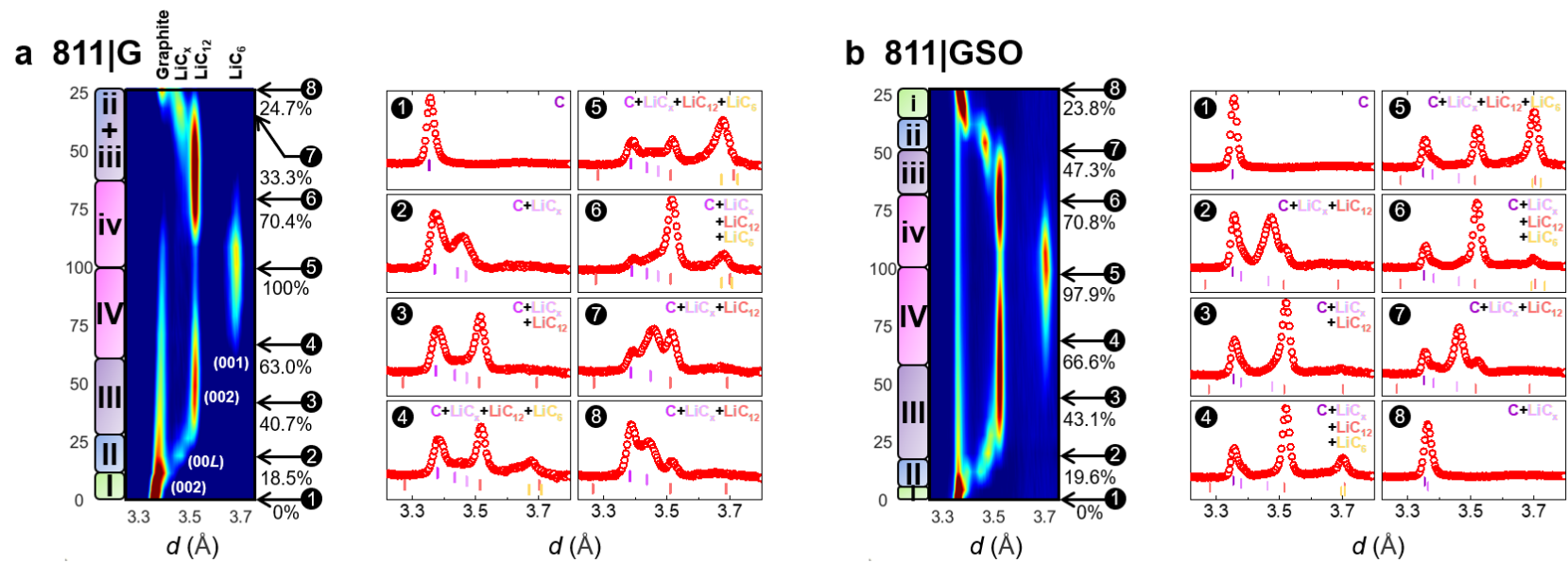

**Fig. S6** *Operando* neutron diffraction contour plots and selected patterns at given SOC of (a) 811|G and (b) 811|GSO cells during the 1$^{st}$ 0.2C cycling over the range from 3.25 to 3.75 Å, where only the (00$L$) reflections of lithiated graphite phases exists. Phases compositions and Miller indices of Bragg reflections are determined and marked on the 811|G contour plot. For each plot, 8 patterns containing phase information are shown on the right, whose SOC could be found through the black circled number markers. Roman numerals on the left of each plot identify main transitions of lithiated graphite during charge/discharge. (I: 2H-graphite → Stage 1L-dilute lithiated graphite; II: Stage 1L-dilute lithiated graphite → Stage 2L/3L/4L-LiC$_x$, x > 12; III: Stage 2L/3L/4L-LiC$_x$, x > 12 → Stage 2-LiC$_{12}$; IV: Stage 2-LiC$_{12}$ → Stage 1-LiC$_6$. Lower-case Roman numerals represent the corresponding reverse transition during discharge.)

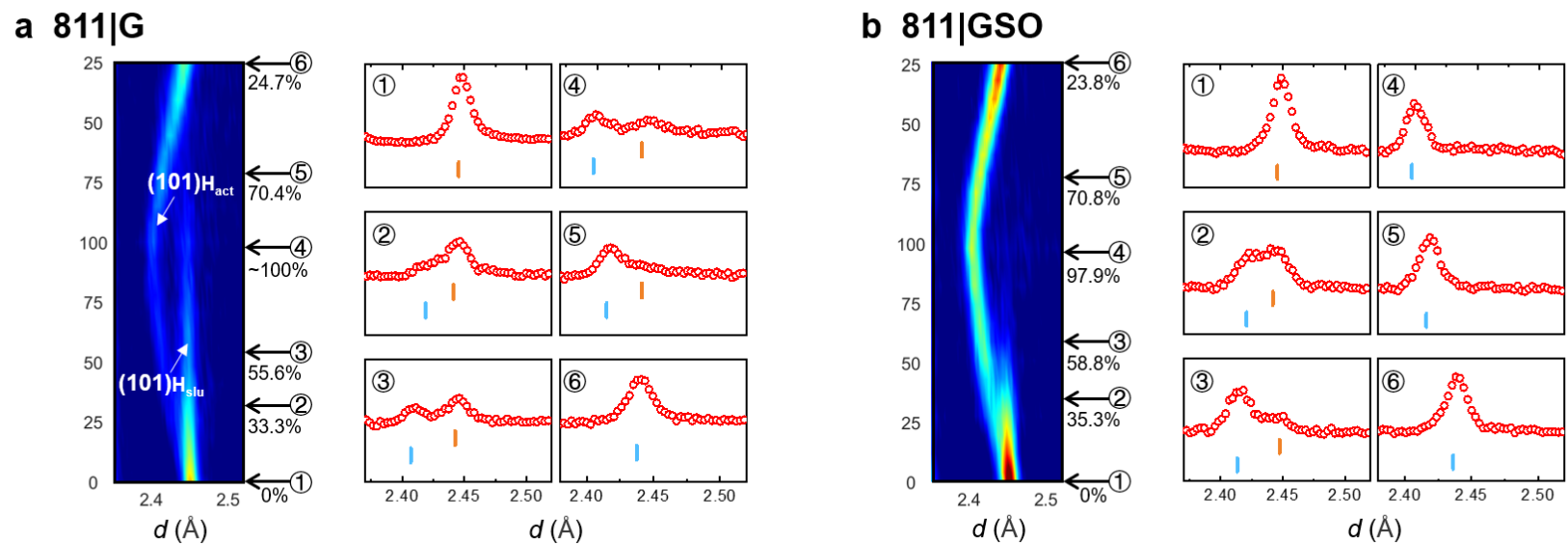

**Fig. S7** *Operando* neutron diffraction contour plots and selected patterns at given SOC of (a) 811|G and (b) 811|GSO cells during the 1st 0.2C cycling over the range from 2.35 to 2.52 Å, where splits of NCM811 (101) reflections are observed. The activate and the sluggish parts of the layered cathode are marked on the 811|G contour plot. For each plot, 6 patterns are shown on the right, whose SOC could be found through the white circled number markers.

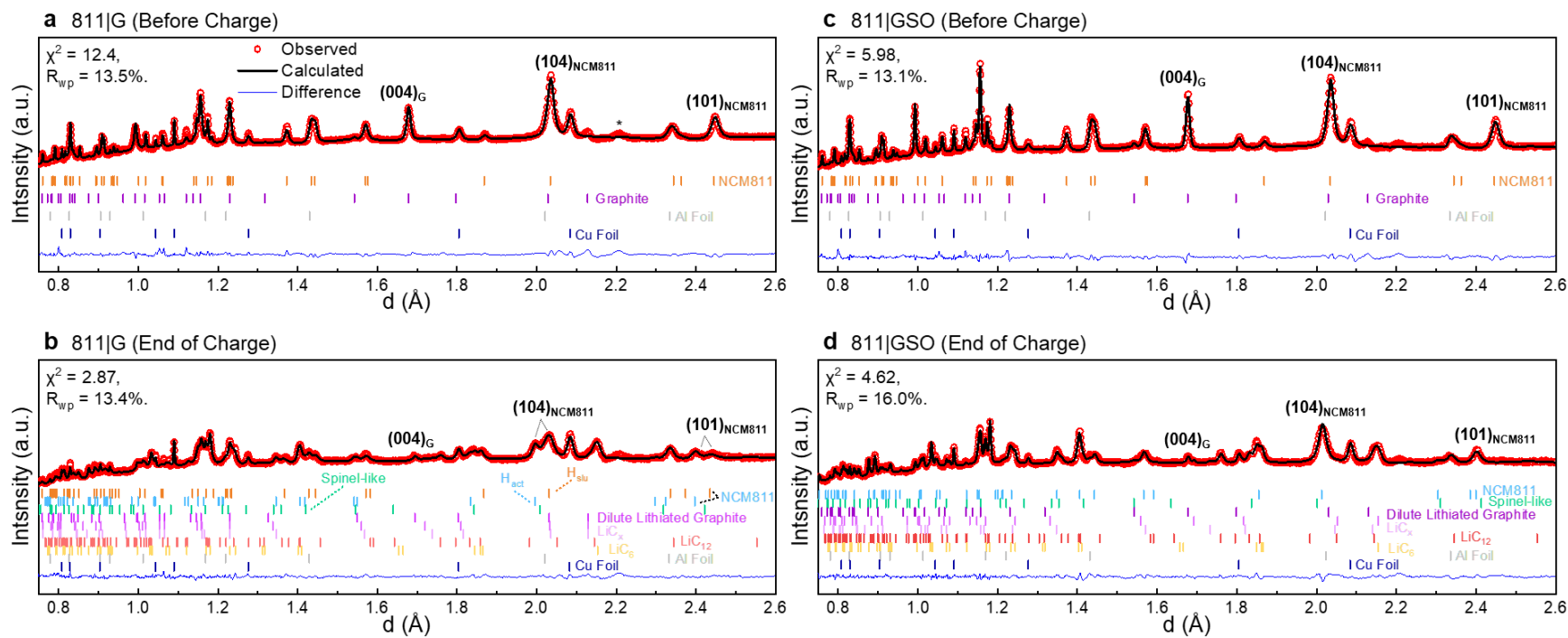

**Fig. S8** Rietveld refinements on *operando* neutron diffraction patterns from 0.75 to 2.6 Å of (a) 811|G before charge, (b) 811|G at the end of charge, (c) 811|GSO before charge, and (d) 811|GSO at the end of charge. Bragg reflections of cathode (including the active $H_{act}$, the sluggish $H_{slu}$, and the spinel-like part), anode (graphite part only), current collectors are identified on the patterns. Other crystalized but low-contribution components like separators of whom the strongest reflection is marked with "*" in (a), or poor-crystallized components like silicon part of anode are ignored in the refinement. Positions of graphite (004) reflections near 1.7 Å, NCM811 (104) reflections near 2.05 Å, and (101) reflections near 2.45 Å are labelled on the patterns.

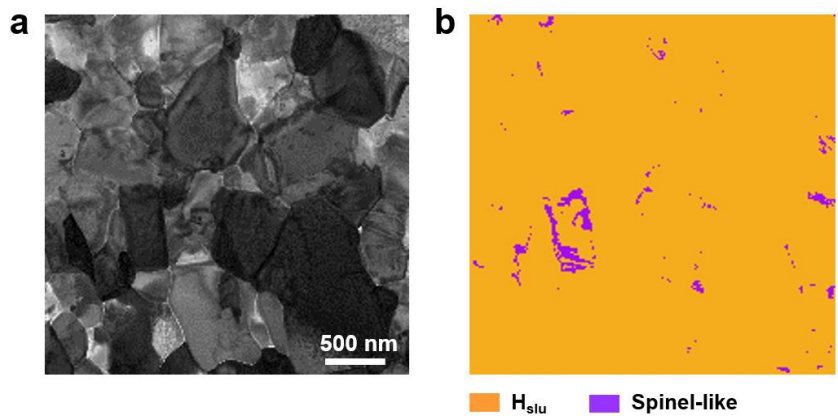

**Fig. S9** (a) Virtual ABF image of the pristine NCM811 primary particles before charge. (b) Phase map for cathode particles shown in (a).

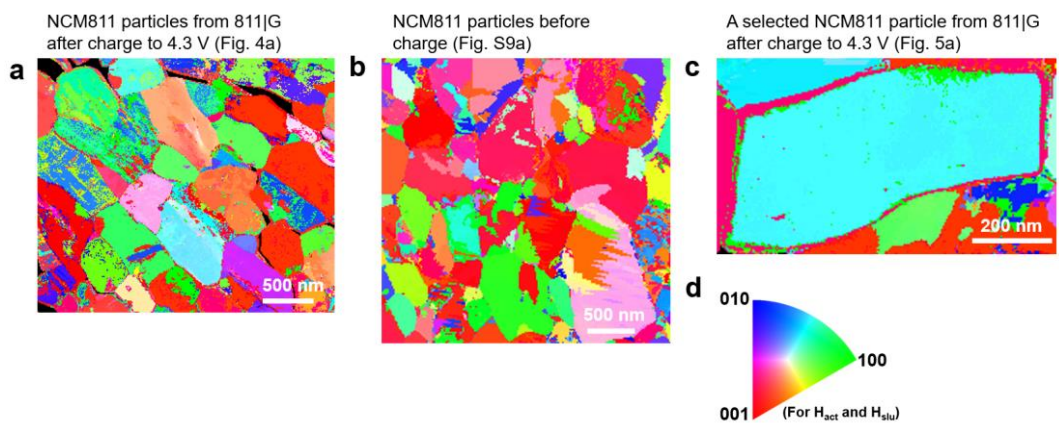

**Fig. S10** Crystallographic orientation maps of cathode particles from (a) 811|G after the 1$^{st}$ charge, (b) particles from pristine NCM811 electrodes before charge, and (c) one particle from charged 811|G at the normal direction (ND) of the observation plane, i.e., the direction along incident electron beams. (d) inverse polar figures for hexagonal phases.

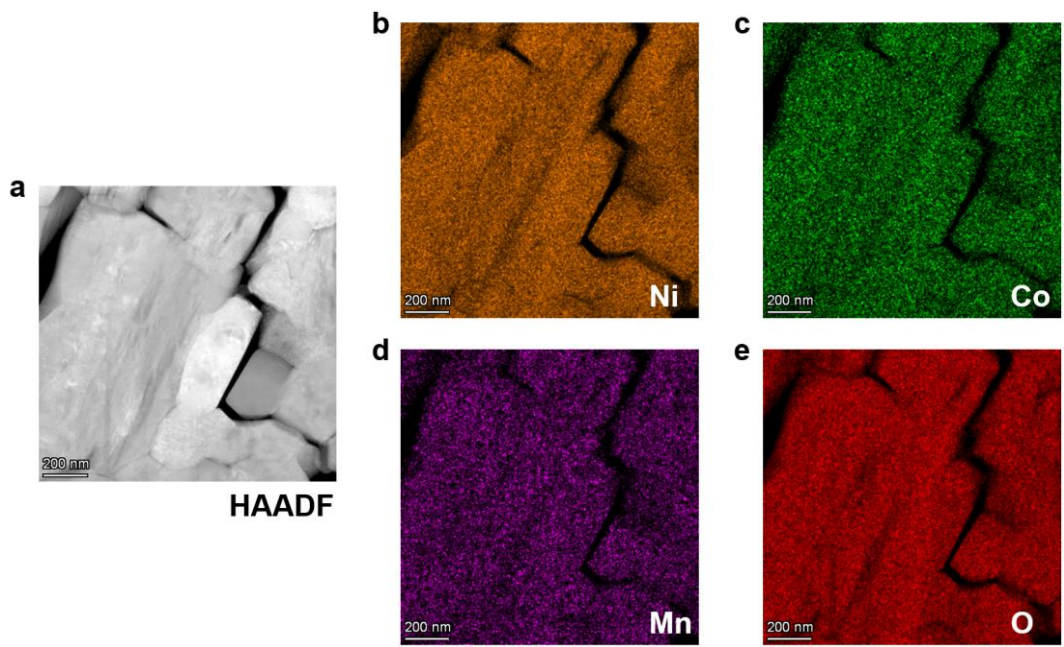

**Fig. S11** (a) HAADF image of NCM811 particles from 811|G after charge and the corresponding EDS mappings of elements including (b) nickel, (c) cobalt, (d) manganese, and (e) oxygen, which exhibit uniform distributions at the delithiation state.

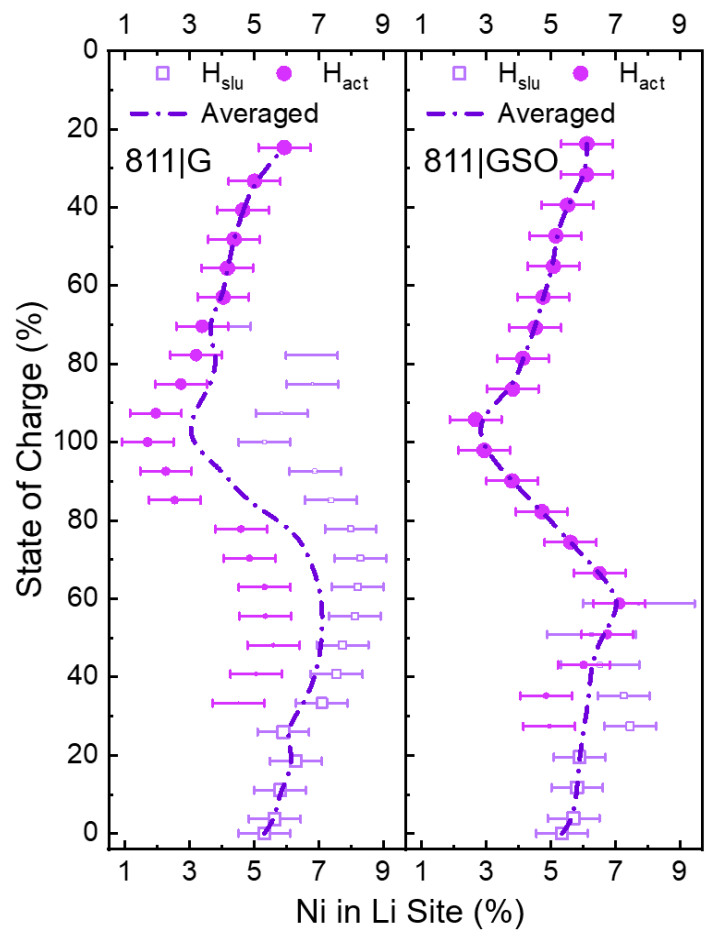

**Fig. S12** Evolution of Li/Ni disordering, represented by the occupancy of Ni in Li site (3*a* site), in layered NCM811 cathode from 811|G (left) and 811|GSO (right) during the 1st 0.2C cycling in the *operando* neutron diffraction experiments.

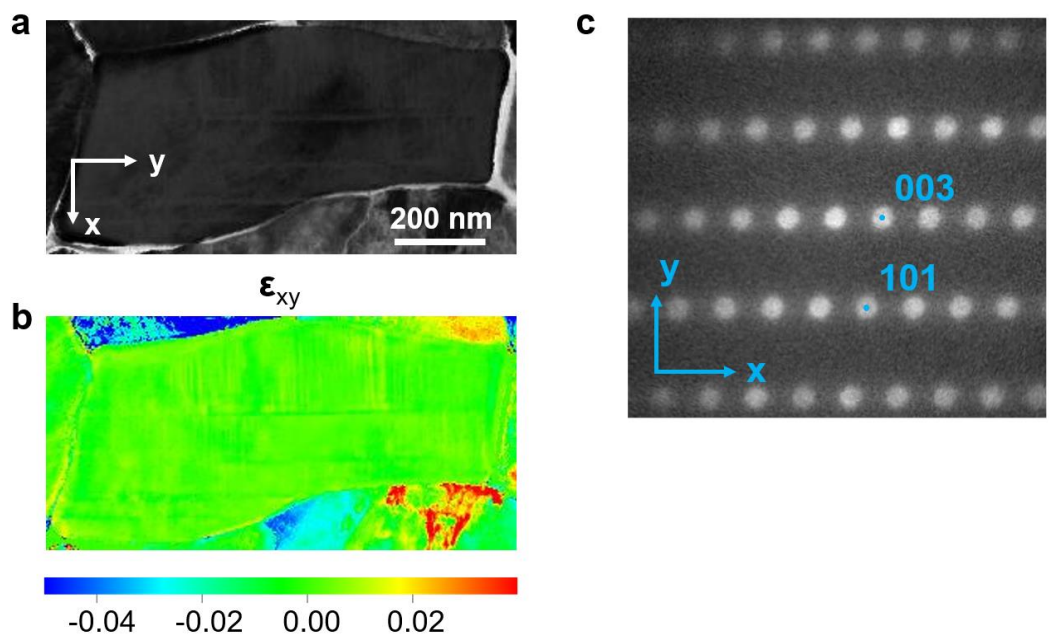

**Fig. S13.** (a) Virtual ABF image of the charged NCM811 particle with intragranular microcracks in Fig. 5a-d. (b) Distribution of the shear strain $\varepsilon_{xy}$ in the selected particle shown in (a). (c) Corresponding summed diffraction pattern.

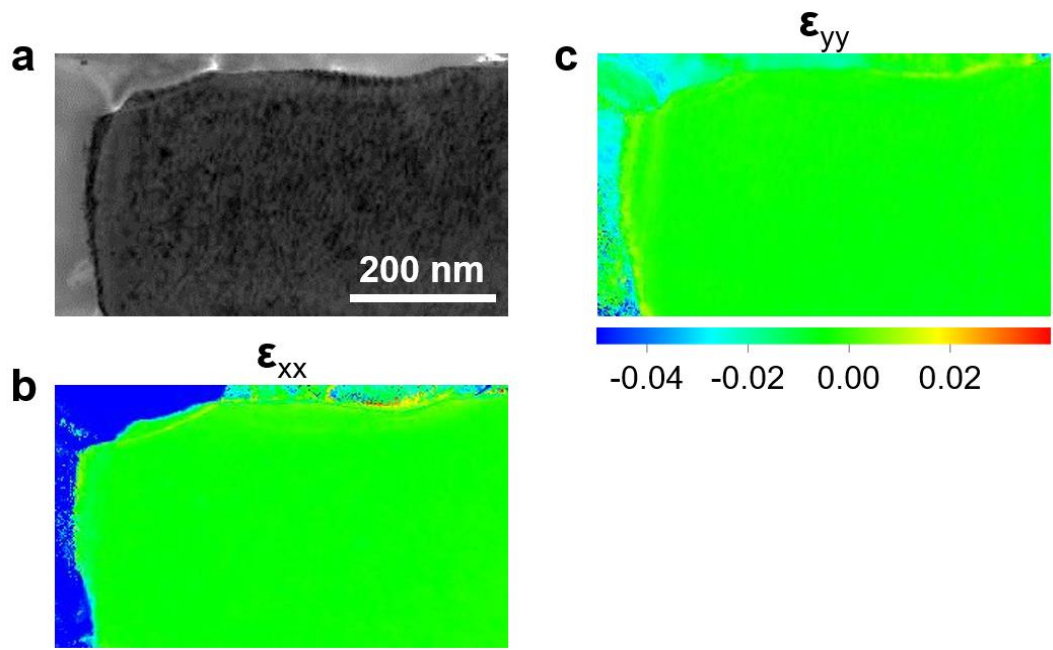

**Fig. S14** (a) Virtual ABF image of a pristine NCM811 particle before charge and strain distribution maps of the selected particle shown in (a), including the normal strain in (b) x-direction, $\varepsilon_{xx}$, and (c) y-direction, $\varepsilon_{yy}$.

**Table S1** Structure models and the initial parameter values for Rietveld refinements of the phases in full cells

| Atom | Wyck. Position | x/a | y/b | z/c | $B_{iso}$ (Å$^2$) | Occ. |
|---|---|---|---|---|---|---|
| NCM811 (Layered Structure) | | | | | | (Cathode) |
| S.G. $R\bar{3}m$ | | | | $a = b = 2.8716$ Å[a], $c = 14.2033$ Å, $\alpha = \beta = 90°, \gamma = 120°$. | | |
| Li | 3a | 0 | 0 | 0 | 1.00 | **1-0.05** |
| Ni (Dis.) | 3a | 0 | 0 | 0 | 1.00 | **0.05** |
| Li (Dis.) | 3b | 0 | 0 | 1/2 | 0.20 | **0.05** |
| Ni | 3b | 0 | 0 | 1/2 | 0.20 | 0.8-**0.05** |
| Co | 3b | 0 | 0 | 1/2 | 0.20 | 0.1 |
| Mn | 3b | 0 | 0 | 1/2 | 0.20 | 0.1 |
| O | 6c | 0 | 0 | **0.24074** | 0.71 | 1 |
| Spinel-like structure due to over-delithiation[b] | | | | | | (Cathode) |
| S.G. $Fd\bar{3}m$ (Origin choice 2) | | | | $a = b = c = 8.0169$ Å, $\alpha = \beta = \gamma = 90°$. | | |
| Li (Tet.) | 8a | 1/8 | 1/8 | 1/8 | 1.00 | 1 |
| Ni (Oct.) | 16d | 1/2 | 1/2 | 1/2 | 0.20 | 0.8 |
| Co (Oct.) | 16d | 1/2 | 1/2 | 1/2 | 0.20 | 0.1 |
| Mn (Oct.) | 16d | 1/2 | 1/2 | 1/2 | 0.20 | 0.1 |
| O | 32e | 0.26523 | 0.26523 | 0.26523 | 0.18 | 1 |
| 2H-Graphite | | | | | | (Anode) |
| S.G. $P6_3/mmc$ | | | | $a = b = 2.4604$ Å, $c = 6.7200$ Å, $\alpha = \beta = 90°, \gamma = 120°$. | | |
| C (1) | 2b | 0 | 0 | 1/4 | 1.50 | 1 |
| C (2) | 2c | 1/3 | 2/3 | 1/4 | 1.50 | 1 |

LiC$_{12}$ (Anode)

*S.G. P6/mmm*      **$a = b = 4.2957$ Å, $c = 7.0417$ Å**, $\alpha = \beta = 90°$, $\gamma = 120°$.

| | | | | | | |
|---|---|---|---|---|---|---|
| Li | 1a | 0 | 0 | 0 | 1.50 | 1 |
| C | 12n | 1/3 | 0 | 0.25000 | 1.50 | 1 |

LiC$_6$ (Anode)

*S.G. P6/mmm*      **$a = b = 4.3118$ Å, $c = 3.6946$ Å**, $\alpha = \beta = 90°$, $\gamma = 120°$.

| | | | | | | |
|---|---|---|---|---|---|---|
| Li | 1a | 0 | 0 | 0 | 1.50 | 1 |
| C | 6k | 1/3 | 0 | 1/2 | 1.50 | 1 |

FCC-Al (Current Collectors)

*S.G. $Fm\bar{3}m$*      $a = b = c = 4.0486$ Å, $\alpha = \beta = \gamma = 90°$.

| | | | | | | |
|---|---|---|---|---|---|---|
| Al | 4a | 0 | 0 | 0 | 0.80 | 1 |

FCC-Cu (Current Collectors)

*S.G. $Fm\bar{3}m$*      $a = b = c = 3.6144$ Å, $\alpha = \beta = \gamma = 90°$.

| | | | | | | |
|---|---|---|---|---|---|---|
| Cu | 4a | 0 | 0 | 0 | 0.80 | 1 |

[a] **Bold parameters** in the table were allowed to vary during refinements;

[b] Minor spinel-like phase was observed in 4D-STEM for high-voltage NCM811. Therefore, spinel-like structure model was taken into consideration only for patterns collected at the end of charge.

**Table S2** Refined cathode structural parameters and anode phase compositions of the 811|G cell under different SOC during the *operando* neutron diffraction experiment

| SOC (%) | Layered NCM811 Cathode State [a] | | | | | | | Graphite Anode State | | | |
|---|---|---|---|---|---|---|---|---|---|---|---|
| | Phase | Stoichiometry of Li [b] | Ni in Li Site (%) | $a$ (Å) | $c$ (Å) | $c/a$ | $V$ (Å$^3$) | C [c] | LiC$_x$ [d] | LiC$_{12}$ | LiC$_6$ |
| - Static - | | | | | | | | | | | |
| 0 | H$_{slu}$ | 1 | 5.3(8) | 2.872(6) | 14.20(3) | 4.95(1) | 101.4(5) | ○ [e] | | | |
| - 1$^{st}$ 0.2C charge - | | | | | | | | | | | |
| 3.70 | H$_{slu}$ | 0.95(4) | 5.6(8) | 2.871(6) | 14.21(3) | 4.95(1) | 101.4(5) | ○ | | | |
| 11.11 | H$_{slu}$ | 0.89(4) | 5.8(8) | 2.869(6) | 14.22(3) | 4.96(1) | 101.4(5) | ○ | ○ | | |
| 18.52 | H$_{slu}$ | 0.85(4) | 6.3(8) | 2.867(6) | 14.23(3) | 4.96(1) | 101.3(5) | ○ | ○ | | |
| 25.93 | H$_{slu}$ | 0.77(4) | 5.9(8) | 2.866(6) | 14.24(3) | 4.97(1) | 101.3(5) | ○ | ○ | | |
| 33.33 | H$_{slu}$ | 0.71(4) | 7.1(8) | 2.866(6) | 14.25(3) | 4.97(1) | 101.4(5) | ○ | ○ | | |
| | H$_{act}$ | 0.68(4) | 4.5(8) | 2.839(6) | 14.32(3) | 5.04(1) | 99.9(4) | | | | |
| 40.74 | H$_{slu}$ | 0.72(4) | 7.6(8) | 2.867(6) | 14.24(3) | 4.97(1) | 101.3(5) | ○ | ○ | ○ | |
| | H$_{act}$ | 0.54(4) | 5.1(8) | 2.833(6) | 14.38(3) | 5.08(1) | 100.0(4) | | | | |
| 48.15 | H$_{slu}$ | 0.72(4) | 7.7(8) | 2.868(6) | 14.23(3) | 4.96(1) | 101.3(5) | ○ | ○ | ○ | |
| | H$_{act}$ | 0.57(4) | 5.6(8) | 2.828(6) | 14.45(3) | 5.11(1) | 100.1(4) | | | | |
| 55.56 | H$_{slu}$ | 0.77(4) | 8.1(8) | 2.868(6) | 14.23(3) | 4.96(1) | 101.4(5) | ○ | ○ | ○ | |
| | H$_{act}$ | 0.53(4) | 5.4(8) | 2.823(6) | 14.46(3) | 5.12(1) | 99.9(4) | | | | |
| 62.96 | H$_{slu}$ | 0.71(4) | 8.2(8) | 2.867(6) | 14.24(3) | 4.97(1) | 101.3(5) | ○ | ○ | ○ | ○ |
| 70.37 | $H_{act}$ | 0.50(4) | 5.3(8) | 2.821(6) | 14.44(3) | 5.12(1) | 99.5(4) | | | | |
| | $H_{slu}$ | 0.72(4) | 8.3(8) | 2.865(6) | 14.24(3) | 4.97(1) | 101.2(5) | ○ | ○ | ○ | ○ |
| | $H_{act}$ | 0.44(4) | 4.9(8) | 2.817(6) | 14.40(3) | 5.11(1) | 99.0(4) | | | | |
| 77.78 | $H_{slu}$ | 0.67(4) | 8.0(8) | 2.862(6) | 14.25(3) | 4.98(1) | 101.1(5) | ○ | ○ | ○ | ○ |
| | $H_{act}$ | 0.43(4) | 4.6(8) | 2.812(6) | 14.31(3) | 5.09(1) | 98.0(4) | | | | |
| 85.19 | $H_{slu}$ | 0.61(4) | 7.4(8) | 2.862(6) | 14.23(3) | 4.97(1) | 101.0(5) | ○ | ○ | ○ | ○ |
| | $H_{act}$ | 0.38(4) | 2.5(8) | 2.814(6) | 13.94(3) | 4.95(1) | 95.6(4) | | | | |
| 92.59 | $H_{slu}$ | 0.52(4) | 6.9(8) | 2.861(6) | 14.24(3) | 4.98(1) | 100.9(5) | ○ | ○ | ○ | ○ |
| | $H_{act}$ | 0.34(4) | 2.3(8) | 2.814(6) | 13.88(3) | 4.93(1) | 95.2(4) | | | | |
| ~100 | $H_{slu}$ | 0.49(4) | 5.3(8) | 2.861(6) | 14.24(3) | 4.98(1) | 101.0(5) | ○ | ○ | ○ | ○ |
| | $H_{act}$ | 0.27(4) | 1.7(8) | 2.813(6) | 13.88(3) | 4.93(1) | 95.2(4) | | | | |
| - 1st 0.2C Discharge - | | | | | | | | | | | |
| 92.59 | $H_{slu}$ | 0.46(4) | 5.9(8) | 2.860(6) | 14.26(3) | 4.99(1) | 101.0(5) | ○ | ○ | ○ | ○ |
| | $H_{act}$ | 0.32(4) | 2.0(8) | 2.814(6) | 13.95(3) | 4.96(1) | 95.7(4) | | | | |
| 85.19 | $H_{slu}$ | 0.45(4) | 6.8(8) | 2.860(6) | 14.27(3) | 4.99(1) | 101.1(5) | ○ | ○ | ○ | ○ |
| | $H_{act}$ | 0.44(4) | 2.7(8) | 2.818(6) | 14.41(3) | 5.11(1) | 99.1(4) | | | | |
| 77.78 | $H_{slu}$ | 0.45(4) | 6.8(8) | 2.861(6) | 14.29(3) | 4.99(1) | 101.3(5) | ○ | ○ | ○ | ○ |
| | $H_{act}$ | 0.46(4) | 3.2(8) | 2.822(6) | 14.46(3) | 5.12(1) | 99.7(4) | | | | |
| 70.37 | $H_{slu}$ | 0.53(4) | 4.1(8) | 2.859(6) | 14.31(3) | 5.00(1) | 101.3(5) | ○ | ○ | ○ | ○ |
| | $H_{act}$ | 0.49(4) | 3.4(8) | 2.824(6) | 14.45(3) | 5.12(1) | 99.8(4) | | | | |

| SOC | Phase | Li occ.[b] | | a (Å) | c (Å) | c/a | V (Å³) | C[c] | LiC$_x$[d] | LiC$_{12}$ | LiC$_6$ |
|---|---|---|---|---|---|---|---|---|---|---|---|
| 62.96 | H$_{slu}$ | —[f] | — | 2.859(6) | 14.31(3) | 5.00(1) | 101.3(5) | ○ | ○ | ○ | ○ |
|  | H$_{act}$ | 0.60(4) | 4.0(8) | 2.835(6) | 14.43(3) | 5.09(1) | 100.4(4) | | | | |
| 55.56 | H$_{slu}$ | — | — | 2.848(6) | 14.34(3) | 5.04(1) | 100.7(5) |  | ○ | ○ | ○ |
|  | H$_{act}$ | 0.64(4) | 4.2(8) | 2.840(6) | 14.39(3) | 5.07(1) | 100.6(4) | | | | |
| 48.15 | H$_{act}$ | 0.68(4) | 4.4(8) | 2.847(6) | 14.35(3) | 5.04(1) | 100.7(5) |  |  | ○ | ○ |
| 40.74 | H$_{act}$ | 0.73(4) | 4.7(8) | 2.852(6) | 14.32(3) | 5.02(1) | 100.8(5) |  |  | ○ | ○ |
| 33.33 | H$_{act}$ | 0.77(4) | 5.0(8) | 2.856(6) | 14.29(3) | 5.00(1) | 100.9(5) | ○ |  | ○ | ○ |
| 24.73 | H$_{act}$ | 0.85(4) | 5.9(8) | 2.861(6) | 14.27(3) | 4.99(1) | 101.2(5) | ○ |  | ○ | ○ |
| End | H$_{act}$ | 0.86(4) | 6.4(8) | 2.863(6) | 14.26(3) | 4.98(1) | 101.2(5) | ○ |  | ○ | ○ |

[a] Information of the slight spinel-like phase observed at high SOC is not included;

[b] Total ratio of Li occupancy at 3$a$ (Li site) and 3$b$ (Ni site) sites. Assume that no Li vacancy existed in the initial structure;

[c] "C" includes pure graphite or dilute lithiated graphite (stage 1L);

[d] "LiC$_x$" includes high-order GICs (stage 2L ~ 4L), e.g., LiC$_{18}$, LiC$_{30}$, LiC$_{40}$, etc.;

[e] ○ - The graphite/GIC phase is observed at this SOC. Blank - The graphite/GIC phase is absent at this SOC;

[f] Due to the low remaining amount of H$_{slu}$, it is difficult to acquire reliable results of the parameters. (Also applicable to **Table S3**)

**Table S3** Refined cathode structural parameters and anode phase compositions of the 811|GSO cell under different SOC during the operando neutron diffraction experiment

| SOC (%) | Layered NCM811 Cathode State | | | | | | | | Graphite Anode State | | | |
|---|---|---|---|---|---|---|---|---|---|---|---|---|
| | Phase | Stoichiometry of Li | Ni in Li Site (%) | $a$ (Å) | $c$ (Å) | $c/a$ | $V$ (Å$^3$) | | C | LiC$_x$ | LiC$_{12}$ | LiC$_6$ |
| | | | | - Static - | | | | | | | | |
| 0 | H$_{slu}$ | 1 | 5.4(8) | 2.872(6) | 14.20(3) | 4.94(1) | 101.4(5) | | ○ | | | |
| | | | | - 1$^{st}$ 0.2C Charge - | | | | | | | | |
| 3.92 | H$_{slu}$ | 0.98(4) | 5.7(8) | 2.870(6) | 14.20(3) | 4.95(1) | 101.3(5) | | ○ | | | |
| 11.75 | H$_{slu}$ | 0.97(4) | 5.8(8) | 2.870(6) | 14.22(3) | 4.96(1) | 101.4(5) | | ○ | ○ | | |
| 19.58 | H$_{slu}$ | 0.92(4) | 5.9(8) | 2.865(6) | 14.23(3) | 4.97(1) | 101.2(5) | | ○ | ○ | ○ | |
| 27.41 | H$_{slu}$ | 1.00(4) | 7.5(8) | 2.868(6) | 14.22(3) | 4.96(1) | 101.3(5) | | ○ | ○ | ○ | |
| | H$_{act}$ | 0.84(4) | 5.0(8) | 2.845(6) | 14.35(3) | 5.04(1) | 100.6(4) | | | | | |
| 35.25 | H$_{slu}$ | 0.95(4) | 7.3(8) | 2.867(6) | 14.23(3) | 4.96(1) | 101.3(5) | | ○ | ○ | ○ | |
| | H$_{act}$ | 0.83(4) | 4.9(8) | 2.841(6) | 14.37(3) | 5.06(1) | 100.4(4) | | | | | |
| 43.08 | H$_{slu}$ | 0.85(8) | 6.5(1.2) | 2.869(6) | 14.24(3) | 4.96(1) | 101.5(5) | | ○ | ○ | ○ | |
| | H$_{act}$ | 0.81(4) | 6.0(8) | 2.839(6) | 14.41(3) | 5.08(1) | 100.5(4) | | | | | |
| 50.91 | H$_{slu}$ | 0.78(9) | 6.3(1.4) | 2.871(6) | 14.25(3) | 4.96(1) | 101.7(5) | | ○ | ○ | ○ | ○ |
| | H$_{act}$ | 0.80(4) | 6.8(8) | 2.836(6) | 14.44(3) | 5.09(1) | 100.6(4) | | | | | |
| 58.75 | H$_{slu}$ | 0.86(11) | 7.7(1.7) | 2.873(6) | 14.22(3) | 4.95(1) | 101.7(5) | | ○ | ○ | ○ | ○ |
| | H$_{act}$ | 0.77(4) | 7.1(8) | 2.831(6) | 14.47(3) | 5.11(1) | 100.5(4) | | | | | |
| 66.58 | $H_{slu}$ | — | — | 2.873(6) | 14.23(3) | 4.95(1) | 101.7(5) | ○ | ○ | ○ | ○ |
| | $H_{act}$ | 0.71(4) | 6.5(8) | 2.828(6) | 14.48(3) | 5.12(1) | 100.3(4) | | | | |
| 74.41 | $H_{slu}$ | — | — | 2.874(6) | 14.23(3) | 4.95(1) | 101.8(5) | ○ | ○ | ○ | ○ |
| | $H_{act}$ | 0.64(4) | 5.6(8) | 2.824(6) | 14.48(3) | 5.13(1) | 100.0(4) | | | | |
| 82.24 | $H_{act}$ | 0.56(4) | 4.7(8) | 2.820(6) | 14.47(3) | 5.13(1) | 99.7(4) | ○ | ○ | ○ | ○ |
| 90.08 | $H_{act}$ | 0.48(4) | 3.8(8) | 2.817(6) | 14.42(3) | 5.12(1) | 99.1(4) | ○ | ○ | ○ | ○ |
| 97.91 | $H_{act}$ | 0.36(4) | 2.9(8) | 2.812(6) | 14.30(3) | 5.08(1) | 98.0(4) | ○ | ○ | ○ | ○ |
| - 1st 0.2C Discharge - | | | | | | | | | | | |
| 94.26 | $H_{act}$ | 0.39(4) | 2.7(8) | 2.814(6) | 14.36(3) | 5.10(1) | 98.4(4) | ○ | ○ | ○ | ○ |
| 86.42 | $H_{act}$ | 0.45(4) | 3.8(8) | 2.819(6) | 14.45(3) | 5.12(1) | 99.4(4) | ○ | ○ | ○ | ○ |
| 78.59 | $H_{act}$ | 0.51(4) | 4.1(8) | 2.822(6) | 14.49(3) | 5.13(1) | 99.9(4) | ○ | ○ | ○ | ○ |
| 70.76 | $H_{act}$ | 0.58(4) | 4.5(8) | 2.825(6) | 14.49(3) | 5.13(1) | 100.2(4) | ○ | ○ | ○ | ○ |
| 62.93 | $H_{act}$ | 0.64(4) | 4.8(8) | 2.830(6) | 14.46(3) | 5.11(1) | 100.3(4) | ○ | ○ | ○ | ○ |
| 55.09 | $H_{act}$ | 0.70(4) | 5.1(8) | 2.837(6) | 14.42(3) | 5.08(1) | 100.5(4) | ○ | ○ | ○ | |
| 47.26 | $H_{act}$ | 0.78(4) | 5.2(8) | 2.842(6) | 14.38(3) | 5.06(1) | 100.6(4) | ○ | ○ | ○ | |
| 39.43 | $H_{act}$ | 0.79(4) | 5.5(8) | 2.849(6) | 14.34(3) | 5.03(1) | 100.9(5) | ○ | ○ | | |
| 31.59 | $H_{act}$ | 0.81(4) | 6.1(8) | 2.855(6) | 14.31(3) | 5.01(1) | 101.0(5) | ○ | ○ | | |
| 23.76 | $H_{act}$ | 0.84(4) | 6.1(8) | 2.860(6) | 14.28(3) | 4.99(1) | 101.1(5) | ○ | ○ | | |
| End | $H_{act}$ | 0.87(4) | 5.9(8) | 2.862(6) | 14.25(3) | 4.98(1) | 101.1(5) | ○ | | | |

**Table S4** Qualitative size and phase distribution of marked primary particles from the 4D-STEM large-area scanning result

| No. | Size | Phase Distribution | | |
|---|---|---|---|---|
| | | $H_{act}$ | $H_{slu}$ | Spinel-like |
| 1 | Medium | Main | Minor near surface, little in the bulk | Little near surface |
| 2 | Large | Main | Minor | Little |
| 3 | Medium | Main | Minor near surface, little in the bulk | Minor near surface, little in the bulk |
| 4 | Medium | Main | Minor | Minor near surface |
| 5 | Small | Almost not observed | Main | Almost not observed |
| 6 | Large | Main | Minor in the bulk | Almost not observed |
| 7 | Small | Main | Little | Little |
| 8 | Large | Minor | Little | Main |
| 9 | Medium | Main | Minor near surface, little in the bulk | Almost not observed |
| 10 | Large | Minor | Main | Almost not observed |
| 11 | Small | Main | Minor near surface, little in the bulk | Almost not observed |
| 12 | Large | Main | Minor near surface, little in the bulk | Little |
| 13 | Medium | Minor | Little | Main |
| 14 | Medium | Little in the bulk | Main | Little in the bulk |
| 15 | Large | Minor | Main | Almost not observed |
| 16 | Small | Main | Minor near surface, little in the bulk | Almost not observed |
| 17 | Medium | Main | Minor near surface, little in the bulk | Almost not observed |

| 18 | Large  | Minor | Main                                    | Almost not observed                     |
| 19 | Large  | Main  | Minor near surface, little in the bulk  | Little near surface                     |
| 20 | Medium | Main  | Little                                  | Minor near surface, little in the bulk  |